\documentclass[manuscript,screen]{acmart}
\AtBeginDocument{%
  }

\setcopyright{acmlicensed}
\copyrightyear{2018}
\acmYear{2018}
\acmDOI{XXXXXXX.XXXXXXX}
\acmISBN{978-1-4503-XXXX-X/18/06}

\usepackage{accsupp}
\usepackage{enumitem}
\usepackage{paralist}
 \usepackage{multirow}
\usepackage{subcaption}
\usepackage{cleveref}
\usepackage{xurl}
\usepackage{multicol}
\usepackage{spverbatim}
\usepackage{fancyvrb,fvextra}
\usepackage{titlesec}
\usepackage{csquotes}

\makeatletter
\def\@subsubsecfont{\sffamily\bfseries\section@raggedright}
\renewcommand\subsubsection{\def\@toclevel{3}%
  \@startsection{subsubsection}{3}{\z@}%
  {-.5\baselineskip \@plus -2\p@ \@minus -.2\p@}%
  {.25\baselineskip}%
  {\ACM@NRadjust{\@subsubsecfont}}}
\renewcommand\paragraph{\def\@toclevel{4}%
  \@startsection{paragraph}{4}{\parindent}%
  {-.3\baselineskip \@plus -2\p@ \@minus -.1\p@}%
  {-3.5\p@}%
  {\ACM@NRadjust{\@parfont\@adddotafter}}}
\makeatother

\newrobustcmd{\SMARTTEXTSC}[1]{\texorpdfstring{\textsc{%
  \BeginAccSupp{ActualText=#1}#1\EndAccSupp{}%
}}{#1}}
\newcommand{\smarttextsc}[1]{\texorpdfstring{\SMARTTEXTSC{#1}}{#1}}

\newcommand{\ie}[0]{\textit{i.e.,}}
\newcommand{\eg}[0]{\textit{e.g.,}}
\newcommand{\etc}[0]{\textit{etc.}}

\newcommand{\etal}[0]{\textit{et al.}}
\newcommand{\vs}[0]{\textit{vs.}}
\newcommand{\wrt}[0]{\textit{w.r.t.}}

\newcommand{\privacypolicy}[1]{%
  \ifvmode %
    \expandafter\MakeUppercase %
  \else%
    \ifnum\spacefactor=3000 %
      \expandafter\MakeUppercase %
    \fi%
  \fi%
  privacy \ifstrequal{#1}{s}{policies}{policy}%
}

\newcommand{\revision}[1]{#1}

\newtheoremstyle{remark}
    {2pt} %
    {2pt} %
    {}          %
    {}          %
    {\itshape} %
    {.}         %
    {.5em}      %
    {}          %

\theoremstyle{remark}
\newtheorem{definition}{\normalfont \bf Definition}[section]

\theoremstyle{remark}
\newtheorem{exmp}{\normalfont \bf Example}[section]

\theoremstyle{remark}
\newtheorem{finding}{Finding}

\theoremstyle{remark}

\newcommand{\tool}[0]{\smarttextsc{PoliGrapher}}
\newcommand{\poligraph}[0]{\smarttextsc{PoliGraph}}

\newcommand{\COLLECT}[0]{\scalebox{.8}[1.0]{\textit{COLLECT}}}
\newcommand{\NOTCOLLECT}[0]{\scalebox{.8}[1.0]{\textit{NOT\_COLLECT}}}
\newcommand{\SUBSUME}[0]{\scalebox{.8}[1.0]{\textit{SUBSUME}}}
\newcommand{\COREF}[0]{\scalebox{.8}[1.0]{\textit{COREF}}}
\newcommand{\PURPOSE}[0]{\scalebox{.8}[1.0]{\textit{PURPOSE}}}

\newcommand{\edge}[1]{\allowbreak$\xrightarrow[]{\text{%
  \raisebox{-1.25pt}[0pt][0pt]{%
    \scalebox{.75}[1.0]{%
      \scriptsize #1%
    }%
  }%
}}$\allowbreak}

\newcommand{\comptt}[1]{\scalebox{.8}[1.0]{\path{#1}}}

\newcommand{\result}[1]{#1}

\newlist{inenum}{enumerate*}{1}
\setlist[inenum]{label=(\arabic*)}

\newcommand{\llmtool}[0]{\smarttextsc{PoliGrapher-LM}}

\begin{document}

\title{PoliGraph: Automated Privacy Policy Analysis using Knowledge Graphs}

\begin{abstract}
Privacy policies disclose how an organization collects and handles personal information. 
Recent work has made progress in leveraging natural language processing (NLP) to automate privacy policy analysis and extract data collection statements from different sentences, considered in isolation from each other.
In this paper, we view and analyze, for the first time, the entire text of a privacy policy in an integrated way.  In terms of methodology:
(1) we define \poligraph{}, a type of knowledge graph that captures statements in a privacy policy as relations between different parts of the text; and 
(2) we revisit the notion of ontologies, previously defined  in heuristic ways, to capture subsumption relations between terms. We make a clear distinction between local and global ontologies to capture the context of individual privacy policies, application domains, and privacy laws.
We develop \tool{}, an NLP tool to automatically extract \poligraph{} from the text using linguistic analysis.
Using a public dataset for evaluation, we show that \tool{} identifies 40\% more collection statements than prior state-of-the-art, with 97\% precision.
In terms of applications, 
\poligraph{} enables automated analysis of a corpus of privacy policies and allows us to: 
(1) reveal common patterns in the texts across different privacy policies, and 
(2) assess the correctness of the terms as defined within a privacy policy.
We also apply \poligraph{} to: 
(3) detect contradictions in a privacy policy, where we show false alarms by prior work, and 
(4) analyze the consistency of privacy policies and network traffic, where we identify significantly more clear disclosures than prior work.
\revision{Finally, leveraging the capabilities of the emerging large language models (LLMs), we also present \llmtool{}, a tool that uses LLM prompting instead of NLP linguistic analysis, to extract \poligraph{} from the privacy policy text, and we show that it further improves coverage.}

\end{abstract}

\author{Hao Cui}
\email{cuih7@uci.edu}
\orcid{0000-0002-7574-2004}
\affiliation{%
  \institution{University of California, Irvine}
  \city{Irvine}
  \state{California}
  \country{USA}
}

\author{Rahmadi Trimananda}
\email{rtrimana@uci.edu}
\orcid{0000-0002-9900-7506}
\affiliation{%
  \institution{University of California, Irvine}
  \city{Irvine}
  \state{California}
  \country{USA}
}

\author{Scott Jordan}
\email{sjordan@uci.edu}
\orcid{0000-0001-5588-311X}
\affiliation{%
  \institution{University of California, Irvine}
  \city{Irvine}
  \state{California}
  \country{USA}
}

\author{Athina Markopoulou}
\email{cuih7@uci.edu}
\orcid{0000-0003-1803-8675}
\affiliation{%
  \institution{University of California, Irvine}
  \city{Irvine}
  \state{California}
  \country{USA}
}



\maketitle

\section{Introduction}
\label{sec:introduction}

\paragraph{Privacy Policies} Privacy laws, such as the General Data Protection Regulation (GDPR)~\cite{gdpr}, the California Consumer Privacy Act (CCPA)~\cite{ccpa}, and other %
data protection laws, require organizations to disclose the personal information they collect, as well as how and why they use and share it. 
Privacy policies are the primary legally-binding way for organizations to disclose their data collection practices to the users of their products. They receive much attention from many stakeholders, such as users who want to exercise their rights,
developers who want their systems to be compliant with privacy laws,
and law enforcement agencies who want to audit organizations' data collection practices and hold them accountable.
Unfortunately, privacy policies are typically lengthy and complicated, making it hard not only for the average user to understand, but also for experts to analyze in depth and at scale~\cite{jordan2021deficiencies}.

\begin{figure*}[t!]
  \centering
  { %
  \small
  \renewcommand{\arraystretch}{0.3}
  \begin{tabular}{|@{} c @{}|@{} c @{}|@{} c @{}|}
    \hline
        & & \\
        \hspace*{0.0mm}
        \includegraphics[width=0.32\textwidth,page=5]{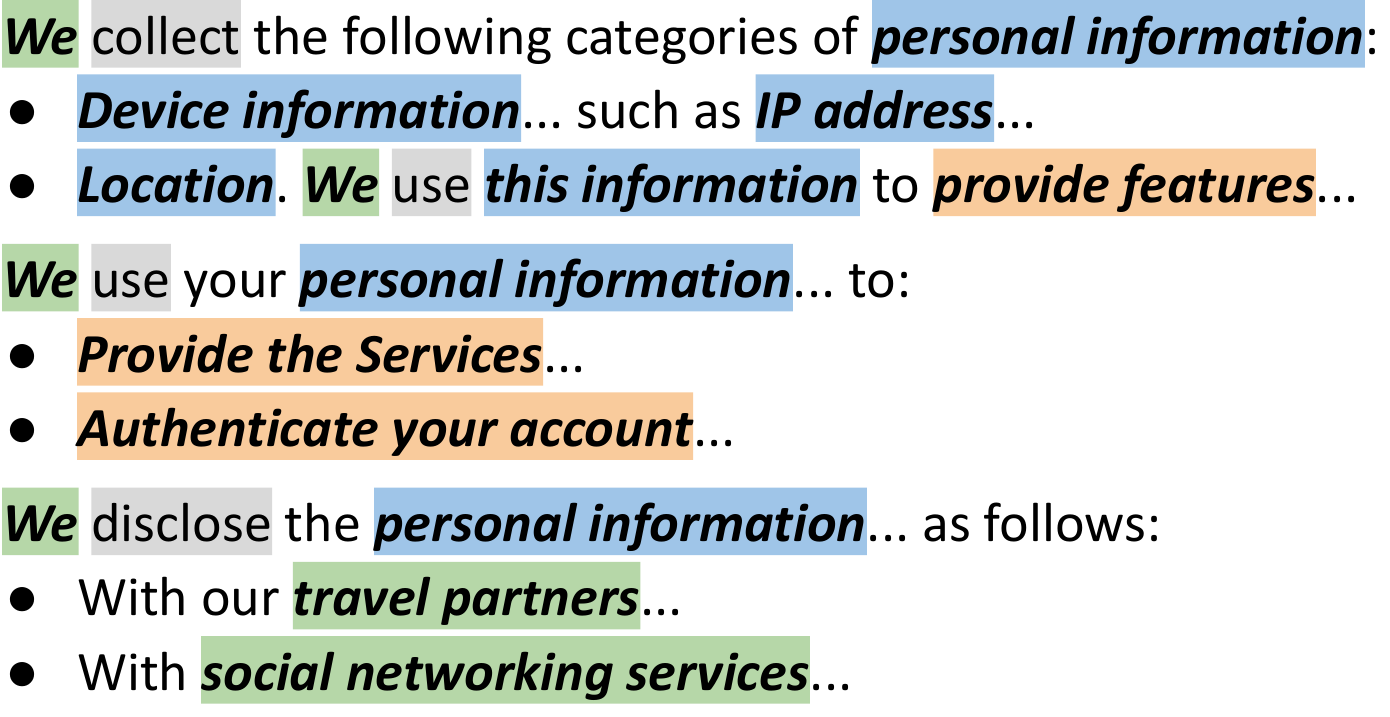}
        & 
        \hspace*{0.0mm}
        \includegraphics[width=0.32\textwidth,page=6]{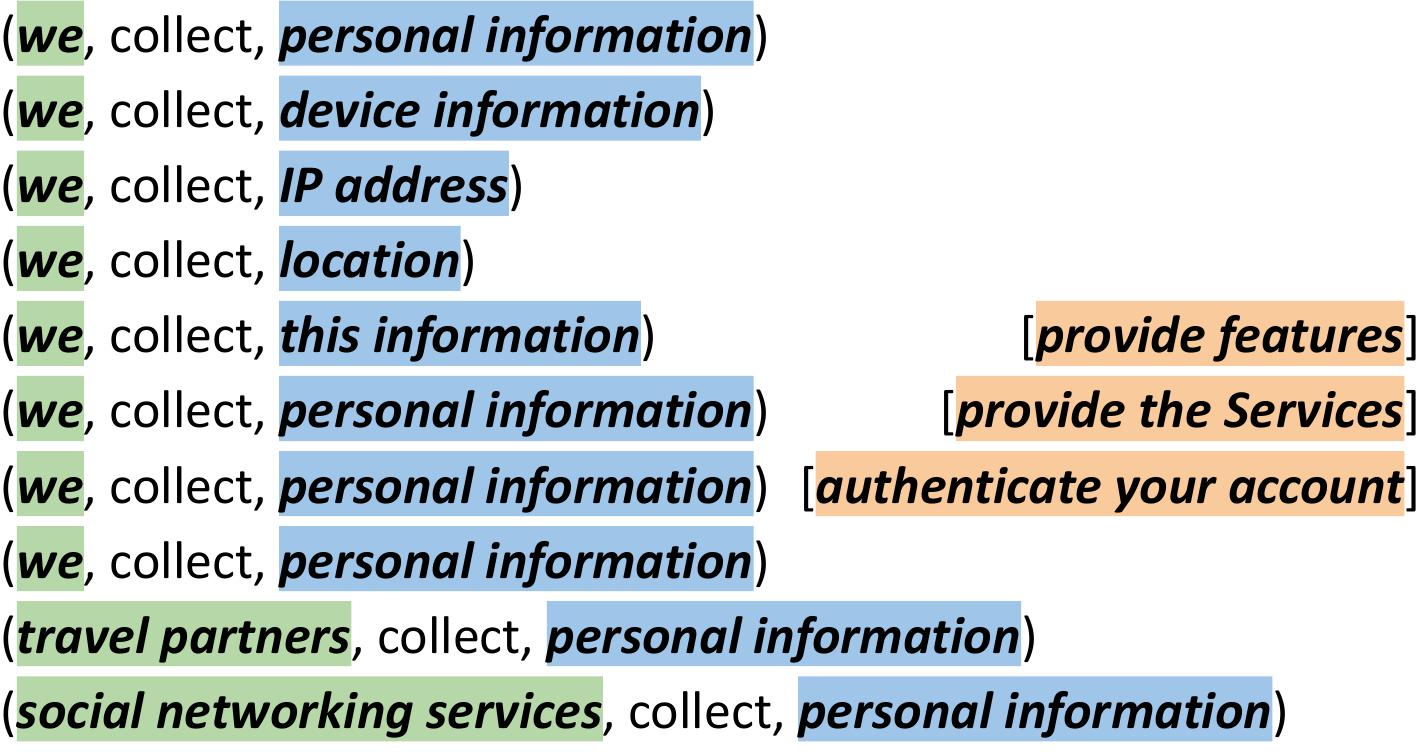}
        \hspace*{0.0mm}
        &
        \hspace*{0.0mm}
        \includegraphics[width=0.335\textwidth,page=7]{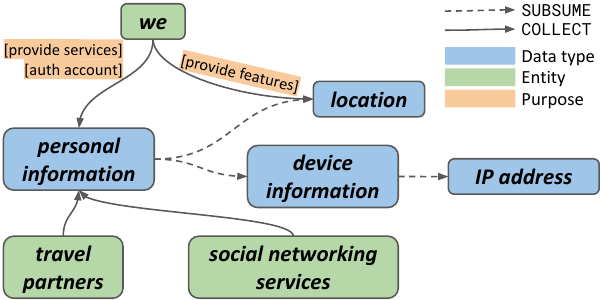}
        \hspace*{0.0mm}
        \\
        & & \\
        \footnotesize{(a)} & \footnotesize{(b)} & \footnotesize{(c)} \\
    \hline
  \end{tabular}
  }
  \caption{
 {\bf Example of a privacy policy and analysis approaches.}  
{\bf (a)} The excerpt is from the policy of KAYAK~\cite{kayak-policy}. %
It contains sections and lists, regarding: what is collected (data type), how it is used (purpose), who receives the information (entity), and references across sentences (\eg{} ``personal information'' relates to other data types; ``this information'' refers to ``location'').
{\bf (b)} Prior work extracts elements found in each sentence, mainly data types and entities, as disconnected tuples. Purposes can also be extracted to extend the tuple~\cite{bui2021consistency,trimananda2022ovrseen}. 
{\bf (c)} \poligraph{} is a knowledge graph that encodes data types, entities, and purposes; and two types of relations between them (collection and subsumption), possibly specified across different sentences. A \COLLECT{} edge represents that a data type is collected by an entity, while edge attributes represent the purposes of that collection. \SUBSUME{} edges represent the subsumption relations between generic and specific terms.}
\label{fig:policy-example}
\end{figure*}

\paragraph{NLP Analysis and Limitations} 
To address this challenge, as well as to facilitate expert analysis~\cite{wilson2016creation} and crowdsourced annotation~\cite{wilson2016crowdsourcing}, the research community has recently applied natural language processing (NLP) to automate the analysis of privacy policies. %
State-of-the-art examples include the following: PolicyLint~\cite{andow2019policylint} extracts data types and entities that collect them, and analyzes potential contradictions within a privacy policy; PoliCheck~\cite{andow2020actions} builds on PolicyLint and further compares the privacy policy statements with the data collection practices observed in the network traffic; Polisis~\cite{harkous2018polisis} and PurPliance~\cite{bui2021consistency} extract data collection purposes; and OVRseen~\cite{trimananda2022ovrseen} leverages PoliCheck and Polisis to associate data types, entities, and purposes.  %
Despite promising results, this body of work also has certain limitations. %

First, existing privacy policy analyzers extract statements (about what is collected, \ie{} data type; who collects it, \ie{} entity; and for what purpose) as disconnected labels~\cite{harkous2018polisis} or 
 tuples~\cite{andow2019policylint,bui2021consistency},
ignoring the links between information disclosed across sentences, paragraphs or sections.
However, today's privacy policies typically have a structure that discloses data types being collected, third-party sharing and usage purposes in separate sections%
\footnote{
We read through 200 privacy policies in our test set (see Section~\ref{sec:evaluation}).
Among them, 135 discuss definitions and practices concerning the same data types in different sections, requiring to put the information together to get the full context about collection, use and sharing of these data types.
In particular, 104 divide content into sections addressing collection, use, and sharing of ``personal information'', resembling the structure shown in Figure~\ref{fig:policy-example}(a).}, as shown in the example in Figure~\ref{fig:policy-example}(a).
Polisis~\cite{harkous2018polisis} uses separate text classifiers to label data types, third-party entities and purposes disclosed in each paragraph. 
Without connecting these labels, it is unclear which data type is collected by which entity, and what purpose applies.
PolicyLint~\cite{andow2019policylint} and PurPliance~\cite{bui2021consistency} adopt tuple representations that put together entities, data types and purposes disclosed in each sentence, as shown in Figure~\ref{fig:policy-example}(b).
However, the tuples still miss context from other sentences. For example, it cannot be inferred from the tuples that the purpose ``provide features'' applies to the collection of ``location''; %
or that the usage purposes and third-party entities in later sections are related to the specific types of ``personal information'' (\eg{} ``device information'') listed in the first section.

Second, because of this incomplete context, prior work %
needs to map and relate the semantics of the terms across different sentences %
by introducing \textit{ontologies} that encode subsumption relations between data types or entities. So far, these ontologies have been built in a manual or semi-automated fashion by domain experts, who define lists of terms commonly found in privacy policy text and other sources (\eg{} network traffic), and subsumption relations between them (\eg{} the term ``device information'' subsumes ``IP address''). The resulting ontologies are not universal: they do not necessarily agree with all privacy policies and need to be adapted to different application domains, \eg{} mobile~\cite{andow2019policylint,andow2020actions,bui2021consistency}, smart speakers~\cite{lentzsch2021hey, echoespaper}, and VR~\cite{trimananda2022ovrseen}.
As a result, they often generate ambiguous or wrong results that require further validation by experts. Manandhar~\etal{}~\cite{manandhar2022smart} recently reported  that state-of-the-art  analyzers~\cite{andow2019policylint,andow2020actions,harkous2018polisis} incorrectly reason about more than half of the privacy policies they analyzed. 

\paragraph{The \poligraph{} Framework} Our key observation is that a policy\footnote{In the rest of the paper, we refer to a privacy policy simply as ``policy''.} should be treated in its entirety, leveraging terms in different sentences that are related. To that end, we make the following methodological contributions.

First, we propose to extract and encode statements in a policy (\ie{} what \textit{data types} are collected, with what \textit{entities} they are shared, and for what \textit{purposes}) into a knowledge graph~\cite{miller1995wordnet,kertkeidkachorn2018t2kg}, which we refer to as \poligraph{}; Figure~\ref{fig:policy-example}(c) shows an example~\footnotemark.
Nodes represent data types or entities. Edges represent relations between nodes, \eg{} an entity may {\em collect} a particular data type, and a more generic data type may {\em subsume} a more specific data type. 
An edge representing data collection may have an attribute indicating the purposes.
The graph in Figure~\ref{fig:policy-example}(c) naturally links the extracted information by merging the same data types and entities and establishing edges between them. It allows inferences such as ``IP address'' being collected for the purpose ``provide services'', and ``location'' being collected by ``travel partners''.

Second, for policies that are not well written, the extracted \poligraph{} may be missing subsumption relations between terms that are not fully defined in the policies.
To supplement the missing relations, we use ontologies, as in prior work~\cite{andow2019policylint,andow2020actions,bui2021consistency}; however, we redefine and use them as follows. First, we consider the subsumption relations extracted from each individual policy as the {\em local ontology} definied by it. Next, we also define additional subsumption relations that encode external knowledge, beyond what is stated in the text of an individual policy; we refer to these as {\em global ontologies}. They can be defined by domain experts, 
using information from multiple policies, or from privacy laws; for example, in Section~\ref{sec:ontology}, we define a data ontology based on the CCPA~\cite{ccpa}. 

\paragraph{\revision{\tool{}: Generating \poligraph{} using Linguistic Analysis}}
We present \tool{}, a methodology and implementation that applies NLP linguistic analysis to automatically extract and build a \poligraph{} from the policy text. To that end, we address several challenges, including coreference resolution, list parsing, phrase normalization, and purpose phrase classification, to extract and link more information than prior work. %
We evaluate \tool{} on a public dataset from PoliCheck%
~\cite{andow2020actions}, consisting of over 6K policies from over 13K mobile apps.
Our manual validation shows that \poligraph{} improves the recall of collection statements from 27\% to 66\%, compared to prior work~\cite{andow2019policylint}, with over 97\% precision.
The improvement is enabled by both the improved NLP techniques and the knowledge graph representation, which can analyze statements spanning multiple sentences and sections \revision{in the policy document.}

\paragraph{Applications} \poligraph{} enables two new types of automated analyses, which were not previously possible.
First, \poligraph{} is used to {\em summarize policies} in our dataset and reveal common patterns across them.
This is possible because \poligraph{}, by representing each policy as a whole, allows inferences about more collection statements.
We find that 64\% of policies disclose the collection of software identifiers and, in particular, cookies. Advertisers and analytics providers are major entities that collect such data. This is further reinforced by the finding that more than half of the policies disclose data usage for non-core purposes, namely for advertising and analytics.
We also find that the use of generic terms for data types (\eg{} ``personal information''), often without more precise definitions, reduces the transparency and leaves the specific data types being collected unknown.
Second, different policies may have different definitions of the same terms.
By clearly separating local ontologies from global ones, \poligraph{} allows us to {\em assess the correctness} of the term definitions.
For example, we find that many policies declare the collected data as ``non-personal information'', which contradicts common knowledge and our CCPA-based global data ontology (see Sections~\ref{sec:ontology} and~\ref{sec:terms-and-definitions}). We also find that non-standard terms are widely used, with varied definitions across policies.

We also apply \poligraph{} to revisit two known applications of policy analysis.
First, to identify {\em contradictions} within a policy, we extend \poligraph{} to analyze negative statements and take into account additional contexts that are crucial for interpreting contradictions, such as (1) fine-grained actions (\eg{} ``sell'' for profit \vs{} ``sharing''), and (2) data subjects (\eg{} children \vs{} general users).
We show that the majority of contradictions found by prior work are false alarms due to language nuances and missing contexts (\eg{} data subjects).
Second, we apply \poligraph{} to analyze {\em data flow-to-policy consistency}. As a result of the improved recall of our approach, we show that prior work~\cite{andow2020actions} has underestimated the number of policies that clearly disclose some sensitive data flows.

\revision{
\paragraph{\llmtool{}: Generating \poligraph{} using LLMs}
The recent developments in Large Langaueg Models (LLMs) have greatly advanced natural language processing. 
To take advantage of and evaluate the capabilities of LLMs for privacy policy analysis, we further develop \llmtool{}, an alternative implementation of \tool{} that extracts \poligraph{} by prompting an LLM.
We address LLMs' limitations, particularly hallucination and coverage errors, by programmatically constraining the output and promting the LLM to reflect on its output. 
Our evaluation shows that \llmtool{} extracts \poligraph{}s with high precision and further improves the recall of collection statements to 83\%, a significant improvement from the linguistic analysis in \tool{}.
However,  the high cost of LLMs is a major barrier to deploy \llmtool{} at scale.
}

\paragraph{Overview} The rest of the paper is structured as follows. 
Section~\ref{sec:related-work} discusses related work.
Section~\ref{sec:knowledge-graph} defines the proposed \poligraph{} framework and the ontologies used with it.
Section~\ref{sec:poligrapher} describes the implementation of \tool{} that uses NLP lintuistic analysis to build \poligraph{} from the text of a policy, and its evaluation.
Section~\ref{sec:applications} presents applications of \poligraph{} to policy analysis.
Section~\ref{sec:llmtool} presents the implementation of \llmtool{} that uses LLMs to build \poligraph{}, and its evaluation.
Finally, Section~\ref{sec:poligraph-discussion} concludes the paper and discusses future directions. \revision{The appendices, uploaded as supplemental materials, provide additional implementation details and evaluation results.}

\section{Related Work}
\label{sec:related-work}

\paragraph{Formalizing Policies} A body of related work focuses on standardizing or formalizing policies. W3C P3P standard~\cite{w3c-p3p} proposed an XML schema to describe policies. The Contextual Integrity (CI)~\cite{nissenbaum2009privacy} framework expresses policies as information flows with parameters including the senders, recipients and subjects of information, data types, and transmission principles that describe the contexts of data collection. None of them replaces text-format policies, but they give insights into defining policies and serve as analysis frameworks. \poligraph{} builds on the CI framework by extracting entities, data types, and part of the transmission principle (\ie{} purposes) from the policy text.

\paragraph{Policy Analysis} Another body of work analyzes policy text. OPP-115~\cite{wilson2016creation} is a policy dataset with manual annotations for fine-grained data practices labeled by experts. Shvartzshnaider \etal{}~\cite{shvartzshnaider2019going}, with the help of crowdsourced workers, analyze CI information flows extracted from policies to identify writing issues, such as incomplete context and vagueness. This manual approach is difficult to scale up for hundreds or thousands of policies due to the significant human efforts.

\paragraph{Automated Policy Analysis} The progress in NLP has made it possible to automate the analysis of unstructured text, such as policy text.
Privee~\cite{zimmeck2014privee} uses binary text classifiers to answer whether a policy specifies certain privacy practices, such as data collection, encryption and ad tracking.
Polisis~\cite{harkous2018polisis}, trained on the OPP-115 dataset, uses 10 multi-label text classifiers to identify data practices, such as the category of data types being discussed and purposes. Classifier-based methods use pre-defined labels which cannot capture the finer-grained semantics in the text.
PolicyLint~\cite{andow2019policylint} first uses NLP linguistic analysis to extract data types and entities in collection statements. PurPliance~\cite{bui2021consistency}, built on top of PolicyLint, further extracts purposes. Conceptually, both works focus on analyzing one sentence at a time, and extracting a tuple \textit{$\langle$entity, collect, data type$\rangle$}, as well as \textit{purpose} in PurPliance, albeit in a separate, nested tuple \textit{$\langle$data type, for / not\_for, $\langle$entity, purpose$\rangle\rangle$}. Unlike \poligraph{}, these  works view extracted tuples individually and do not infer data practices disclosed across multiple sentences.

\paragraph{Knowledge Graphs} Graphs are routinely used to integrate knowledge bases as relationships between terms~\cite{miller1995wordnet}. Google has used a knowledge graph built from crawled data to show suggestions in search results~\cite{google-knowledge-graph}. OpenIE~\cite{angeli2015leveraging} and T2KG~\cite{kertkeidkachorn2018t2kg} use NLP to build knowledge graphs from a large corpus of unstructured text. In \poligraph{}, we use knowledge graphs, for the first time, to represent policies.

\par\smallskip
\revision{The \poligraph{} framework first appeared in~\citet{usenix-version}. Compared to that, this journal submission provides additional results and new materials. In particular, the design and evaluation of \llmtool{} in~\Cref{sec:llmtool} is new to this paper, and was motivated by the breakthroughs of large language models (LLMs) that happened after the acceptance of the original paper to the USENIX Security Symposium.}

\section{The \poligraph~Framework}
\label{sec:knowledge-graph}

In this section, we introduce \poligraph{}, our proposed representation of the entire text of a policy as a knowledge graph.
We also revisit the related notion of ontologies, %
and we propose a new definition and use it with \poligraph{}.

\subsection{Defining \poligraph}
\label{sec:terminology}

We define \poligraph{} as a knowledge graph that captures statements in a policy considered as a whole. Throughout this section, we will use Figure~\ref{fig:policy-example} as our running example to illustrate the terminology and definitions.

Privacy laws, such as the GDPR~\cite{gdpr} and the CCPA~\cite{ccpa}, require that organizations disclose their practices regarding data collection, sharing and use in their policies. To capture these three aspects of disclosures in the policy, we represent the corresponding three kinds of terms in \poligraph{}: what \textit{data types} are collected, with what \textit{entities} they are shared, and  for what \textit{purposes} they are used.

\begin{compactitem}[$\bullet$]
\item \emph{Data type:} This kind of terms refers to the type of data being collected.  In Figure~\ref{fig:policy-example}(a), ``location'' is a specific collected data type. Generic terms can be used as well, \eg{} ``personal information'' and ``device information''.
\item \emph{Entity:} This kind of terms refers to the organization that receives the collected data. It can be the first party if it is the developer of the product (\eg{} website, mobile app, \etc{}) that writes the policy, namely ``we'' in Figure~\ref{fig:policy-example}(a); or, otherwise, a third party such as ``travel partners'' in Figure~\ref{fig:policy-example}(a).
\item \emph{Purpose:} Policies may also specify purposes.\footnote{In this paper, we refer to \textit{purposes} of processing of personal data as specified in the GDPR, namely the purposes of collection, use, and sharing. US laws often distinguish among the three, \eg{} the CCPA appears to require a policy to separately disclose the purposes of collection / use and the purposes of sharing personal information.}
In Figure~\ref{fig:policy-example}(a), purposes include ``provide services'',  ``authenticate your account'', and ``provide features''.
\end{compactitem}
\vspace{4pt}

In \poligraph{}, we represent data types and entities as two different types of nodes. Furthermore, we encode the following relations between them as edges.
\begin{compactitem}[$\bullet$]
    \item \emph{\COLLECT{} edge}: An entity $n$ may collect a  data type $d$. 
    In Figure~\ref{fig:policy-example}(a), ``personal information'' is collected by the first-party entity ``we'', but it is also shared with the entity ``travel partners'' (a third party).
    More formally, a \COLLECT{} edge $e_c=n$\edge{COLLECT}$d$ between an entity $n$ and a data type $d$  represents that $d$ is collected by $n$, namely $collect(n, d)$. %
    
    \item \emph{\SUBSUME{} edge}: A generic term (\textit{hypernym}) may subsume a more specific term (\textit{hyponym}). For example, ``personal information'' subsumes ``device information'' and ``location'', and ``device information'' in turn subsumes ``IP address''. More formally, a \SUBSUME{} edge {\small$e_s=hyper$\edge{SUBSUME}$hypo$} connects nodes $hyper$ and $hypo$, where both nodes $hyper,~hypo$ are data types or both are entities, and it represents that the more generic term $hyper$ subsumes the more specific term $hypo$, namely $subsume(hyper,~hypo)$.

    \item \emph{Purposes as edge attributes}:
    We represent purposes by assigning them as a list of attributes {\small$Purposes(e_c)\!=\!\{p_1, p_2, ...\}$} to each \COLLECT{} edge $e_c$. This is a natural choice that fits how policies are written: one or more purposes are typically associated with a data type and an entity.  In Figure~\ref{fig:policy-example}(a), entity ``we'' (\ie{} KAYAK) collects ``this information'', which refers to ``location'', for the purpose ``to provide features''.
    The purpose ``to provide features'' is captured in the list of attributes $Purposes(e_c)\!=\!\{\textit{provide features}\}$ assigned to the \COLLECT{} edge $e_c\!=\!we$\edge{COLLECT}$location$.
\end{compactitem}
\vspace{1pt}

In summary, we define \poligraph{}, representing knowledge about data collection, sharing and use disclosed within a particular policy, as follows.

\begin{definition}{\textbf{\poligraph{}.}}
\label{def:poligraph}
A \poligraph{} $G = \langle D, N;$ $E_{S}, E_{C}; P \rangle$ is a directed acyclic graph.
Each node represents a term that is either a data type $d \in D$ or an entity $n \in N$. Each edge can be either a \SUBSUME{} edge $e_s \in E_{S}$, or a \COLLECT{} edge $e_c \in E_{C}$ as defined above.
A \COLLECT{} edge $e_c$ has a list of attributes \textit{Purposes}$(e_c)=\{p_1, p_2, ...\}$, where $p_i \in P$.
\end{definition}

Figure~\ref{fig:policy-example}(c) shows the \poligraph{} representation of the policy text in Figure~\ref{fig:policy-example}(a). The technical details about building the graph from verbatim text, such as how to map the co-reference term ``this information'' to ``location'', are provided in Section~\ref{sec:implementation}. 
Next, we define relations that can be inferred from \poligraph{} about policy text.

\begin{definition}{\textbf{\textit{Subsumption Relation.}}}
In a \poligraph{} $G$, we say that a term $t_1$  (hypernym) \textit{subsumes} another term $t_2$ (hyponym), denoted as $subsume(t_1, t_2)$, \textit{iff} there exists a path from $t_1$ to $t_2$ in $G$ where every edge is a \SUBSUME{} edge.\footnote{A subsumption relation is naturally transitive. To simplify other definitions, we also make it reflexive, \ie{} every term subsumes itself.
}
\end{definition}

\begin{definition}{\textbf{\textit{Collection Relation.}}}
\label{def:data-collection}
In a \poligraph{} $G$, we say an entity $n\!\in\!N$ \textit{collects} a data type $d\!\in\!D$, denoted as $\mathit{collect}(n, d)$, \textit{iff} there exists an entity $n' \in N$ and a data type $d' \in D$ where
$subsume(n', n) \wedge{} subsume(d', d)$\footnote{That is, a policy may disclose data collection using generic terms. For instance, in Figure~\ref{fig:policy-example}(c), we have $\mathit{collect}(\textit{we}, \textit{IP address})$ because ``IP address'' is also ``personal information''.}
and the edge $n'$\edge{COLLECT}$d'$ exists in $G$.
\end{definition}

\begin{definition}{\textbf{\textit{Set of Purposes.}}}
\label{def:set-of-purposes}
Following Definition~\ref{def:data-collection}, if a purpose $p \in \mathit{Purposes}(n'$\edge{COLLECT}$d')$, we say $n$ collects $d$ for the purpose $p$. We denote the set of all instances of such $p$ in $G$ as a set $\mathit{purposes}(n, d)$. 
\end{definition}

Beyond what is captured by individual nodes, edges, and attributes, the strength of \poligraph{} is that it allows us to make inferences.
In Figure~\ref{fig:policy-example}(c), there is no direct edge from ``travel partners'' to ``location'', but we can still infer that ``location'' may be shared with ``travel partners'' and ``social network services''.   
Furthermore, %
we can also infer that $collect$(\textit{we}, \textit{location}) and $purposes$(\textit{we}, \textit{location}) = $\{ \textit{provide features} \}$. 
Such data practices that are implied, but not explicitly stated, would be missed by prior work that only processes individual sentences, and possibly by human readers as well.

Prior state-of-the-art work would have extracted a list of tuples, as depicted in the example of Figure~\ref{fig:policy-example}(b). PolicyLint ~\cite{andow2019policylint}  and follow-up works~\cite{andow2020actions,lentzsch2021hey} extract 2-tuples: \textit{$\langle$entity, data type$\rangle$}. Purposes can be extracted independently 
and appended to form a longer 3-tuple \textit{$\langle$entity, data type, purpose$\rangle$} as in OVRseen~\cite{trimananda2022ovrseen}, or put in a nested tuple as in PurPliance~\cite{bui2021consistency}. %
In all cases, those tuples are extracted from individual sentences that are disconnected from each other. As a result, prior work would fail to identify implied statements. In contrast,  \poligraph{} connects terms with the same semantics in different sentences, allowing
inferences and improving coverage.

Another major strength of \poligraph{} is that its modular design makes it easy to extend to capture additional relations.
In Section~\ref{sec:negative-statements}, we present \poligraph{} extensions to handle finer-grained semantics, including negative edges and subtypes of \COLLECT{} edges to distinguish among data actions (\eg{} ``sell'' for profit \vs{} ``sharing''), as well as data subjects.

\subsection{Ontologies}
\label{sec:ontology}

Policies refer to data types and entities at different semantic  granularities. For example, ``device information'' in Figure~\ref{fig:policy-example}(a) is a generic data type that subsumes ``IP address'' and maybe other more specific data types.
Prior work~\cite{andow2019policylint,andow2020actions,trimananda2022ovrseen} has introduced hierarchies of terms, namely ontologies, to define the subsumption relations between data types or entities.
They typically define the data and entity ontologies heuristically and manually, by considering a combination of information found in the network traffic and in the policy text, as well as using domain expertise to organize terms into hierarchies.

We revisit the notion of ontologies under the \poligraph{} framework.
First, \poligraph{} naturally captures subsumption relations described in an individual policy, which form the \textit{local ontology}.
Ideally, if a policy is written in a clear and complete way, it should either use specific terms, or clearly define generic terms that will be captured by the corresponding local ontology.
In practice, policies are not perfectly written and parts of the ontology may be missing.
For example, in Figure~\ref{fig:policy-example}(a), the term ``social networking services'' is not further explained.
Furthermore, some policies may provide misleading definitions, \eg{} ``geolocation'' is described as non-personal information, whereas it is widely considered personal by the public and privacy laws (see Section~\ref{sec:terms-and-definitions}).
Second, we define and design \textit{global ontologies} that encode external knowledge or ground truth, as in prior work.
For the first time, the distinction between local and global ontologies provides a principled way to summarize an individual policy, as well as to assess the completeness and correctness of definitions by comparing the local against the global ontologies.

\subsubsection{Local Ontologies}
\label{sec:local-ontology}

In \poligraph{}, \SUBSUME{} edges between data types or entities induce a directed acyclic graph, which we refer to as a local ontology, capturing the relations between more generic and more specific terms, as defined within a particular policy. We define local data and entity ontologies as follows.

\begin{definition}{\textbf{\textit{Local Ontology.}}}
\label{def:local-ontology}
A local ontology is either a data ontology $o_d=\langle D, E_{d} \rangle$ or an entity ontology $o_n=\langle N, E_{n} \rangle$, a directed acyclic graph that is a subgraph of \poligraph{} $G = \langle D, N; E_{S}, E_{C}; P \rangle$, in which every node is a data type $d \in D$ or an entity $n \in N$, and every edge $e_d \in E_{d}, e_n \in E_{n}$ (where $E_{d},E_{n} \subset E_{S}$) is a \SUBSUME{} edge. 
\end{definition}

In Figure~\ref{fig:policy-example}(c), the four blue nodes containing data types form the local data ontology: the root node is ``personal information'' and the leaf nodes are ``location'' and ``IP address''.
The local entity ontology, which contains the three green nodes, does not have a nontrivial hierarchical structure because the policy does not further explain the terms “travel partners” and “social networking services”.

\subsubsection{Global Ontologies}

We define a global ontology to encode external knowledge, \ie{} outside a particular policy, which we consider as ground truth in that context. 
It provides a reference against which we can compare and evaluate individual policies, as well as a complement to missing definitions in policies.

\begin{figure}[t!]
\centering
\begin{minipage}{.6\linewidth}
    \includegraphics[width=\columnwidth]{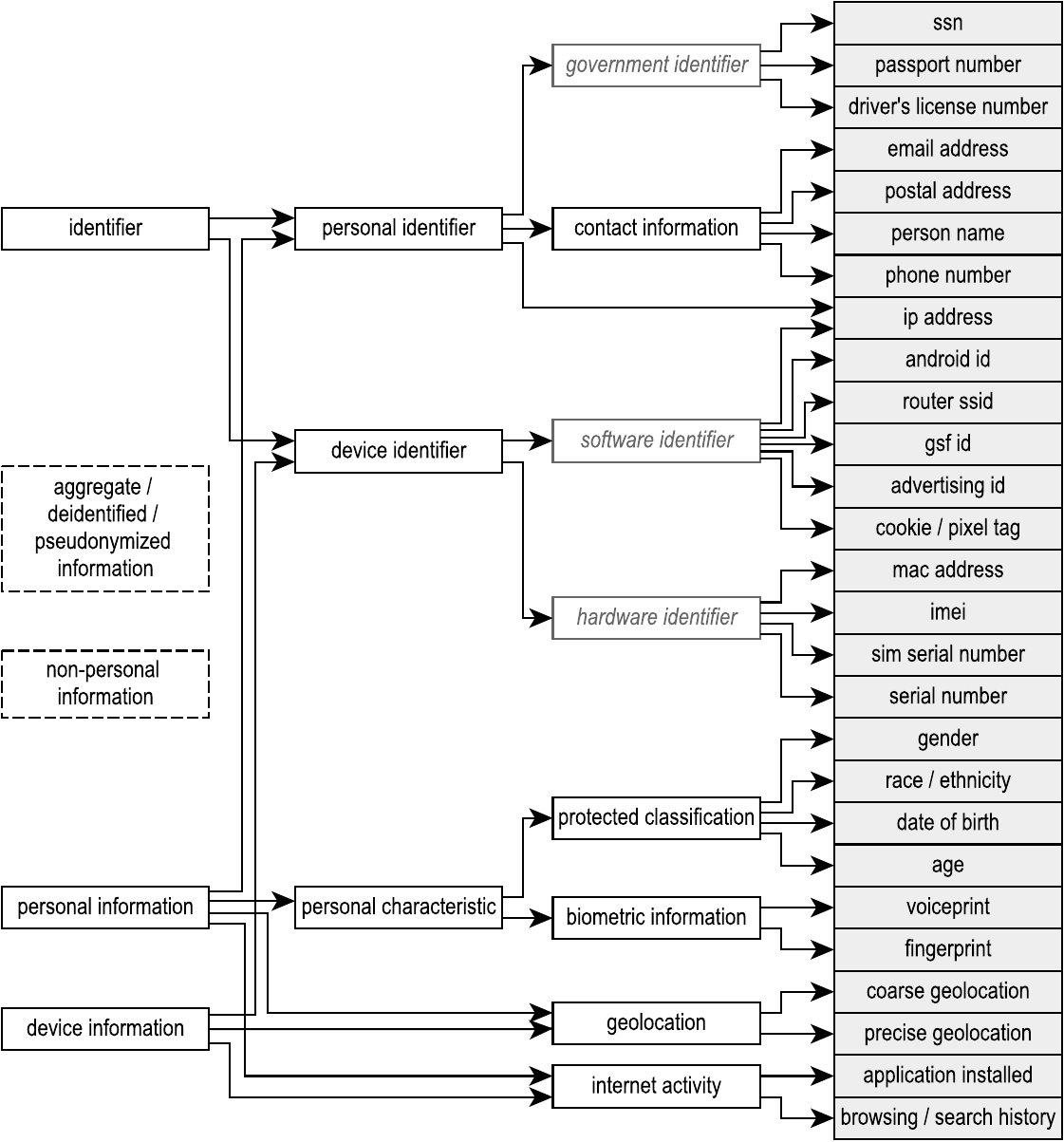}
    \caption{Global Data Ontology based on the CCPA.}
    \label{fig:data-ontology}
\end{minipage}\hfill
\begin{minipage}{.36\linewidth}
    \includegraphics[width=\columnwidth]{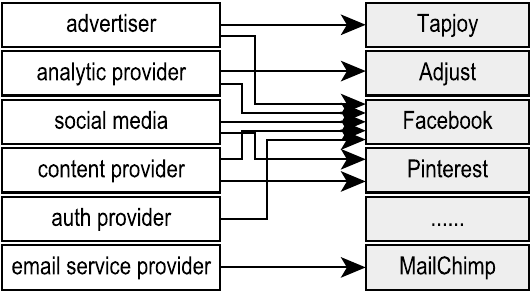}
    \caption{Global Entity Ontology based on~\cite{tracker-radar,crunchbase-data}.}
    \label{fig:entity-ontology}  
\end{minipage}
\end{figure}

\begin{definition}{\textbf{\textit{Global Ontology.}}}
\label{def:global-ontology}
A global ontology is either a data ontology $O_d=\langle D_{d},E_{d} \rangle$ or an entity ontology $O_n=\langle N_{n},E_{n} \rangle$ that is a directed acyclic graph, where every node is a data type $d \in D_{d}$ or an entity $n \in N_{n}$, and every edge $e_d \in E_{d}$ or $e \in E_{n}$ is a \SUBSUME{} edge. %
\end{definition}

Prior work~\cite{andow2019policylint,andow2020actions,trimananda2022ovrseen} has implicitly and heuristically defined such global ontologies, by taking into account and combining
the union of all subsumption relations extracted from policies in their corpus, and
the data types and entities observed in the actual system's output (\eg{} network traffic).
However, such global ontologies have not been universal: they may include subjective judgment, and they typically do not apply across application domains. %
For example, PoliCheck's data ontology does not assume ``personal information'' to include ``device information'': this contradicts the content of the policy depicted in Figure~\ref{fig:policy-example}(a). Although we recognize that there is no single way to define perfect global ontologies, we propose that we rely on authoritative sources, such as privacy laws, to define them. An example is described next, but other designs can be used with \poligraph{} as well.

\paragraph{Global Data Ontology Based on the CCPA}
As a concrete, illustrative example, we propose a global data ontology that is based on the CCPA~\cite{ccpa}. %
The CCPA governs the collection, use, and sharing of personal information, as defined therein, by companies that do business in California.
To build the CCPA-based global data ontology, we start with the definition of ``personal information'' in CCPA Section 1798.140(v)(1), which includes, but is not limited to, specific data types, including a person's name, social security number, postal address, email address, and IP address. We place such specific data types into the ontology as leaf nodes. Then, since policies often disclose the collection of \textit{categories} of these specific data types, \eg{} ``contact information'' instead of ``email address'' and ``postal address'', we organize these specific data types into categories delineated by subsumption relations. The CCPA's definition of personal information also includes categories for which it does not list specific data types, \eg{} ``biometric information''. In such cases, we include the categories in the global data ontology and augment it with common specific data types, \eg{} ``biometric information'' includes ``voiceprint'' and ``fingerprint''. Similarly, the CCPA uses the term ``device identifier'' but does not define it, while we include it as a category in the global data ontology, and place specific device identifiers in that category. Figure~\ref{fig:data-ontology} shows the CCPA-based global data ontology.
The above is meant as a concrete example of a global ontology based on a privacy law. Different laws (\eg{} GDPR) can lead to different global ontologies.

\paragraph{Global Entity Ontology}
Privacy laws give examples of the types of entities, but not the exhaustive list of entities, with whom an organization may share personal information.
We follow policies that often categorize entities by service types.
We obtain a list of entities and their categories from the DuckDuckGo Tracker Radar dataset~\cite{tracker-radar} and a CrunchBase-based dataset~\cite{crunchbase-data}. Based on these sources, containing 4,709 entities in total, we propose a simple two-level ontology that classifies entities into six categories as shown in Figure~\ref{fig:entity-ontology}.

The global ontologies, serving as the ground truth of subsumption relations, are used to categorize unorganized data types and entities in \poligraph{}s (see Section~\ref{sec:summarization}), assess the correctness of term definitions in individual policies (see Section~\ref{sec:terms-and-definitions}), as well as complement term definitions in case of missing definitions when we check vague disclosures of data flows (see Section~\ref{sec:flow-to-policy-consistency-analysis}).

\section{\tool{}: Extracting \poligraph{} through Linguistic Analysis}
\label{sec:poligrapher}

In this section, we present \tool{}, the NLP-based system that we implement to generate \poligraph{} from the text of a policy, and its evaluation.

\subsection{Implementation}
\label{sec:implementation}

\begin{figure}[t!]
	\centering
    \includegraphics[width=\columnwidth,page=1]{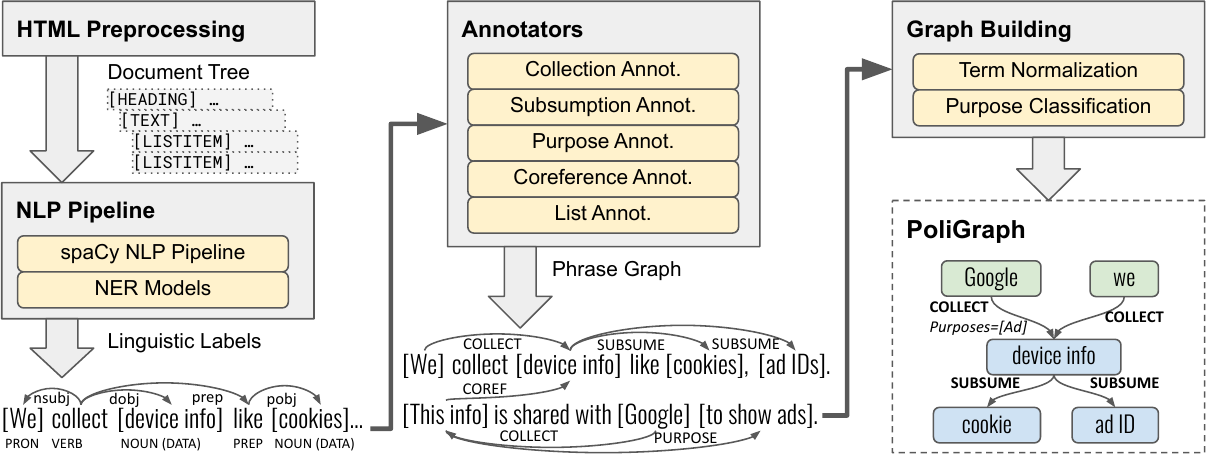}
	\caption{\textbf{Overview of \tool{} implementation.} First, \tool{} preprocesses the HTML document to produce a simplified document tree structure. Second, the NLP pipeline takes the document tree and labels sentences with linguistic labels. Third, the labeled sentences are annotated by the annotators to produce a phrase graph containing all the annotations. Finally, the graph building stage deploys term normalization and purpose classification to transform the phrase graph into a \poligraph{}.}
	\label{fig:tool-overview}
\end{figure}

\tool{} use NLP linguistic analysis to identify phrases that represent data types, entities, and purposes in the privacy policy, and extract relations between these phrases.
\Cref{fig:tool-overview} gives an overview of \tool{} implementation.

\subsubsection{NLP on Structured Documents}
\label{sec:nlp-on-structured-document}
\privacypolicy{s} are usually published online as structured documents, mainly in HTML format, while NLP models expect plain text input. Simply stripping HTML tags, such as headings and lists, would result in a loss of semantics.
As the first step, \tool{} preprocesses an HTML document to a simplified tree structure which preserves some important document structures.
The document tree is used to generate complete sentences as input for the NLP pipeline.

\paragraph{HTML Preprocessing}
Policies are usually published online as structured documents, mainly in HTML format, while NLP models expect plain text input. Simply stripping HTML tags, such as headings and lists, would result in a loss of semantics.
As the first step, \tool{} preprocesses each HTML document to a simplified document tree which preserves three important document structures: \textit{heading}, \textit{list item}, and general \textit{text}.
The document tree helps to generate complete sentences as input for NLP.
Please see~\Cref{appendix:html-preprocessing} for details.

\paragraph{NLP Pipeline}
\tool{} is built based on the spaCy library~\cite{spacy3} and its RoBERTa-based NLP pipeline~\cite{liu2019roberta,spacy-trf-model}.
The NLP pipeline labels text with linguistic labels originating from English linguistic features, including word lemmas, part-of-speech, sentence segmentation, and syntactic dependency trees (see the output of ``NLP Pipeline'' in Figure~\ref{fig:tool-overview}). These features are syntactic and thus require no domain adaptation.

To identify data types and entities in a policy, \tool{} uses named entity recognition (NER), a standard NLP technique to classify noun phrases into a given set of labels. In our case, we use two labels: \textit{DATA} for data types and \textit{ENTITY} for entities. To train the NER model, we use a synthetic training set that combines generated sentences and real policy text prelabeled by an existing NER model and rule-based NER.
The model achieves 96.1\% precision and 89.4\% recall on our test set.
Please see~\Cref{appendix:ner-methodology} for details.

\subsubsection{Annotators}
\label{sec:annotators}

\begin{table}[t!]
    \centering
    \caption{Overview of annotators in \tool{}.}{}
    \label{tab:annotator-overview}
    \small{
\newcommand{\YellowBox}[1]{\setlength{\fboxsep}{1pt}\colorbox{yellow}{#1}}
\newcommand{\LimeBox}[1]{\setlength{\fboxsep}{1pt}\colorbox{lime}{#1}}
\newcommand{\GreyBox}[1]{\setlength{\fboxsep}{1pt}\colorbox{gray!20}{#1}}

\begin{tabular}{@{}ll@{}}
\toprule
\textbf{Annotator} & \textbf{Example} (based on the policy in  Figure~\ref{fig:policy-example}(a)) \\ \midrule

\begin{tabular}[c]{@{}l@{}}Collection Annotator\end{tabular} &
\begin{tabular}[c]{@{}l@{}}
\YellowBox{Entity}\edge{COLLECT}\LimeBox{Data} \\
\textit{\eg{} \YellowBox{We} collect ... \LimeBox{personal information}}
\end{tabular}
\\ \midrule

\begin{tabular}[c]{@{}l@{}}Subsumption Annotator\end{tabular} &
\begin{tabular}[c]{@{}l@{}}
\YellowBox{Hypernym}\edge{SUBSUME}\LimeBox{Hyponym} \\
\textit{\eg{} \YellowBox{Device information}... such as \LimeBox{IP address}...}
\end{tabular}
\\ \midrule

\begin{tabular}[c]{@{}l@{}}Purpose Annotator\end{tabular} &
\begin{tabular}[c]{@{}l@{}}
\YellowBox{Data}\edge{PURPOSE}\LimeBox{Purpose} \\
\textit{\eg{} We use your \YellowBox{personal information} ... to:} \\
\textit{\phantom{\eg{}}\LimeBox{Provide the Services}...}
\end{tabular}
\\ \midrule

\begin{tabular}[c]{@{}l@{}}Coreference Annotator\end{tabular} &
\begin{tabular}[c]{@{}l@{}}
\YellowBox{Reference}\edge{COREF}\LimeBox{Main mention} \\
\textit{\eg{} We collect ... \LimeBox{personal information}: ...} \\
\textit{\phantom{\eg{}} We use \YellowBox{this information} to ...}
\end{tabular}
\\ \midrule

List Annotator &
\begin{tabular}[c]{@{}l@{}}
\YellowBox{Preceding sentence}\edge{SUBSUME / COLLECT}\LimeBox{List item} \\
\textit{\eg{} \GreyBox{\rm TEXT} We collect ... \colorbox{yellow}{following information}:} \\
\textit{\phantom{\eg{}} ~~~~\GreyBox{\rm LISTITEM} - \LimeBox{Device information}...} \\
\textit{\phantom{\eg{}} ~~~~\GreyBox{\rm LISTITEM} - \LimeBox{Location}...}
\end{tabular}
\\
\bottomrule
\end{tabular}
}

\end{table}

In \tool{}, we refer to the modules that identify relations between phrases\footnotemark{} as \textit{annotators}. The relations are stored as edges in a graph structure, which we call a \textit{phrase graph}. The phrase graph is still an intermediate step, in which phrases referring to the same thing have not been merged.

\footnotetext{We use ``phrase'' to refer to verbatim words and phrases in the \privacypolicy{} text. We use ``term'', which appears in previous sections as well, to refer to the normalized forms (see \Cref{sec:building-knowledge-graph}) of phrases that appear in \poligraph{}s.}

To extract relations between phrases, annotators search for phrases matching specific syntactic patterns~\cite{de-marneffe-etal-2014-universal}.
In contrast to prior work that hardcodes heuristics to perform the search~\cite{andow2019policylint}, we use \textit{dependency matching} to specify desired patterns as configurable rules~\cite{spacy-rules}. For example, in the collection annotator, we set the rule \comptt{collect|gather|obtain|...:ROOT} to match the root verb ``collect''. Then the sub-rule \comptt{ENTITY:subj} matches the subject ``we'' under the verb as the entity, and another sub-rule \comptt{DATA:obj} matches the object ``device information'' as the data type. The annotator then adds a \COLLECT{} edge between them in the phrase graph.

By dividing linguistic analysis tasks into five annotators, each of them focuses on a specific set of patterns. \Cref{tab:annotator-overview} outlines the patterns and relations that each annotator tries to identify. Note that some edge types (\COREF{} and \PURPOSE{}) exist only in the phrase graph and will be converted in the final \poligraph{}. We discuss each annotator as follows.

\paragraph{Collection Annotator}
The collection annotator finds affirmative sentences that disclose data collection, use or sharing, extracts entities and data types, and adds \COLLECT{} edges from entities to data types in these sentences.
The annotator matches around 40 verbs and 20 sets of syntactic patterns. \Cref{tab:collection-patterns} lists some of the patterns in the active voice. For clarity, we do not list patterns in the passive voice (\eg{} ``this information is shared with...'') and composite patterns (\eg{} ``allow us to collect...''), but they are all handled by the annotator. We gather these patterns from actual \privacypolicy{s} in the dataset which we use to evaluate \poligraph{} (see \Cref{sec:applications}).

The collection annotator only labels \COLLECT{} edges for affirmative statements. To distinguish affirmative sentences from negative and interrogative ones, it checks the existence of negative modifiers (\eg{} \textit{not}, \textit{never}) and interrogative words in the dependency tree. While negative statements are by default excluded, we extend \poligraph{} in \Cref{sec:negative-statements} to use the information to analyze negative statements.

\paragraph{Subsumption Annotator}
The annotator identifies subsumption relations between phrases and adds \SUBSUME{} edges from a hypernym to its hyponyms.
It matches 11 syntactic patterns of subsumption as shown in \Cref{tab:subsumption-patterns}. These extend the patterns used in prior work~\cite{andow2019policylint,andow2020actions,bui2021consistency}.

\paragraph{Purpose Annotator}
The annotator identifies phrases that describe purposes of data collection in three forms:
\begin{inenum}
    \item \textit{in order to $\langle$verb$\rangle$ ...};
    \item \textit{to $\langle$verb$\rangle$ ...};
    \item \textit{for ... purpose(s)}.
\end{inenum}
\looseness=-1
It links such purpose phrases to corresponding data types with \PURPOSE{} edges, which are not part of \poligraph{} and will be converted into $Purposes(\cdot)$ attributes on the corresponding \COLLECT{} edges in \poligraph{}. For example, in the sentence ``We use this information \textit{to provide ads}'', the purpose phrase ``to provide ads'' is linked to the data type ``this information''.

\begin{table}[t!]
\begin{minipage}{.49\linewidth}
    \caption{Syntactic patterns used by the collection annotator.}
    \label{tab:collection-patterns}
    \centering
    \resizebox{\columnwidth}{!}{\newcommand{\BAdjust}[0]{\setlength{\fboxsep}{1pt}\colorbox{yellow}{Google}}
\newcommand{\BWe}[0]{\setlength{\fboxsep}{1pt}\colorbox{yellow}{We}}
\newcommand{\BDeviceID}[0]{\setlength{\fboxsep}{1pt}\colorbox{lime}{device IDs}}

\small
\begin{tabular}{@{}lll@{}}
\toprule
\begin{tabular}[c]{@{}l@{}}\textbf{Root Verbs}\\
\textit{(Examples: \textit{\setlength{\fboxsep}{1pt}\colorbox{yellow}{ENTITY}\edge{COLLECT}\colorbox{lime}{DATA})}}
\end{tabular}
& \textbf{Syntatic Patterns}
\\ \midrule

\begin{tabular}[c]{@{}l@{}}share, trade, exchange, disclose\\
\textit{(\BWe{} share your \BDeviceID{} with \BAdjust{}.)}
\end{tabular}
&
\begin{tabular}[c]{@{}l@{}}ENTITY:nsubj\\ DATA:dobj\\ with,ENTITY:pobj\end{tabular}
\\ \midrule

\begin{tabular}[c]{@{}l@{}}
\scalebox{.9}[1.0]{collect, gather, obtain, get, receive, solicit, acquire, request}\\
\textit{(\BAdjust{} may collect your \BDeviceID{}.)}
\end{tabular}
&
\begin{tabular}[c]{@{}l@{}}ENTITY:nsubj\\ DATA:dobj\end{tabular}
\\ \midrule

\begin{tabular}[c]{@{}l@{}}provide, supply\\
\textit{(\BWe{} provide \BAdjust{} with your \BDeviceID{}.)}
\end{tabular}
&
\begin{tabular}[c]{@{}l@{}}ENTITY:nsubj\\ ENTITY:dobj\\ with,DATA:pobj\end{tabular}
\\ \midrule

\begin{tabular}[c]{@{}l@{}}provide, supply, release, disclose, transfer, transmit,\\ sell, rent, lease, give, pass, divulge, submit\\
\textit{(\BWe{} may transmit \BDeviceID{} to \BAdjust{}.)}
\end{tabular}
&
\begin{tabular}[c]{@{}l@{}}ENTITY:nsubj\\ DATA:dobj\\ to,ENTITY:pobj\end{tabular}
\\ \midrule

\begin{tabular}[c]{@{}l@{}}use, keep, access, analyze, process, store, save, hold,\\log, utilize, record, retain, preserve, need, maintain\\
\textit{(\BAdjust{} may use your \BDeviceID{}.)}
\end{tabular}
&
\begin{tabular}[c]{@{}l@{}}ENTITY:nsubj\\ DATA:dobj\end{tabular}
\\ \midrule

\begin{tabular}[c]{@{}l@{}}have, get, gain (access to)\\
\textit{(\BAdjust{} has access to your \BDeviceID{}.)}
\end{tabular}
&
\begin{tabular}[c]{@{}l@{}}ENTITY:nsubj\\ access,to,DATA:pobj\end{tabular}
\\ \midrule

\begin{tabular}[c]{@{}l@{}}make (use of)\\
\textit{(\BAdjust{} makes use of \BDeviceID{}.)}
\end{tabular}
&
\begin{tabular}[c]{@{}l@{}}ENTITY:nsubj\\ use:dobj\\ of,DATA:pobj\end{tabular}
\\ %

\bottomrule
\end{tabular}
}
\end{minipage}\hfill
\begin{minipage}{.49\linewidth}
    \caption{Syntactic patterns used by the subsumption annotator.}
    \label{tab:subsumption-patterns}
    \centering
    \resizebox{\columnwidth}{!}{\small{
\begin{tabular}{@{}ll@{}}
\toprule
\textbf{Phrases}                     & \textbf{Sentences} \\ \midrule
$X$ such as $Y_1, Y_2...$            & $X$ includes $Y_1, Y_2...$ \\
such $X$ as $Y_1, Y_2...$            & $X$ includes but is not limited to $Y_1, Y_2...$ \\
$X$, for example, $Y_1, Y_2...$      & \\
$X$, e.g. / i.e. $Y_1, Y_2...$       & \\
$X$, which includes $Y_1, Y_2...$    & \\
$X$ including / like $Y_1, Y_2...$     & \\
\multicolumn{2}{@{}l@{}}{$X$, especially / particularly, $Y_1, Y_2...$} \\
\multicolumn{2}{@{}l@{}}{$X$, including but not limited to, $Y_1, Y_2...$} \\
$Y_1, Y_2...$ (collectively $X$)  \\
\bottomrule
\multicolumn{2}{@{}l}{\scriptsize\begin{tabular}[c]{@{}l@{}}
$X$ = hypernym phrase; $Y_1, Y_2...$ = hyponym phrases.
\end{tabular}}
\end{tabular}
}
}   
\end{minipage}
\end{table}

\paragraph{Coreference Annotator}
The annotator resolves pronouns (\eg{} ``it'', ``they'', \etc) or phrases modified by demonstrative determiners (\eg{} ``this'', ``those'', \etc) to the phrases which they refer to. This task, known as \textit{coreference resolution}, is a non-trivial NLP task.
Prior work~\cite{andow2019policylint,andow2020actions,bui2021consistency} could not handle coreferences properly, which has resulted in a loss of semantics and misinterpretation of many collection statements.

We find that existing coreference resolution models~\cite{joshi-etal-2019-bert,coreferee} cannot handle non-personal references well, whereas they are commonly found in \privacypolicy{s}.
To address the issue, we design a heuristic-based coreference annotator that handles common forms of coreferences in \privacypolicy{s}.
First, for a phrase starting with a determiner ``this'', ``that'', ``these'', ``those'' or ``such'' (\eg{} ``these providers''), the annotator looks backward for the nearest phrase with the same root word (\eg{} ``ad providers'') in the same or previous sentence. Specifically, if the root word is ``data'' or ``information'' (\eg{} ``this information''), the annotator looks backward for the nearest data type labeled by NER as the referent.
Second, for a pronoun like ``it'', ``this'', ``they'', or ``these'', the annotator tries to infer whether the pronoun refers to a data type or an entity based on existing \SUBSUME{} edges, and looks backward for the nearest data type or entity.
The annotator links coreference phrases to the referred phrases with \COREF{} edges, which are only used in phrase graphs to resolve coreferences.

We evaluate our method on 200 coreferences from our test set (see \Cref{sec:evaluation}). 168 are resolved correctly. Four coreferences are partially resolved because they refer to multiple phrases while the annotator supports only one referent for each phrase. The other 28 are not resolved or are resolved wrongly. This yields 84-86\% accuracy, which suggests that it outperforms general-purpose coreference models.
For example, Coreferee \cite{coreferee} reported 82-83\% accuracy on general corpora. However, as it does not treat data types as named entities, it cannot resolve coreferences of data types.

\paragraph{List Annotator}
This special annotator uses the document tree to discover relations between list items and their preceding sentence.
First, if a noun phrase modified by ``following'' or ``below'' (\eg{} ``the following information'') precedes list items, it adds \SUBSUME{} edges from the phrase to list items.
Second, it propagates relations between the preceding sentence of a list and \textit{any} list item to \textit{all} list items in case other annotators fail to label them.

\subsubsection{From Phrase Graph to \poligraph}
\label{sec:building-knowledge-graph}

The final step of \tool{} is to build a \poligraph{} from a phrase graph. This involves merging phrases (data types and entities) with the same meaning to one node and converting purpose phrases to edge attributes.

\paragraph{Normalizing Data Types and Entities}
\tool{} starts by mapping data types and entities in the phrase graph to their normalized forms. For example, ``contact details'' and ``contact data'' are synonyms to the normalized term ``contact information'' which we want to keep in the \poligraph{}.

For data types and entities in our global ontologies (see \Cref{sec:ontology}), we consider them as standard terms and write regular expressions to capture their synonyms.
For example, the regular expression \comptt{contact\b.*\b(information|data|detail|method)} matches synonyms of ``contact information''. \tool{} maps these synonyms to ``contact information'', which aligns with the term in the global data ontology, as the normalized form.
We also programmatically create regular expressions for variants of company names from public datasets. If a phrase is not a standard term and thus does not match any regular expressions, \tool{} simply strips stop words and takes the lemmatized form of the phrase as the normalized form. For example, ``your vehicle records'' is normalized to ``vehicle record''. This is usually enough to capture variants of the same term caused by word inflections.

For coreferences, as the annotator has linked each coreference phrase to what it refers to,
\tool{} follows the \COREF{} edge to find the referred phrase and use the same normalized form. For example, in \Cref{fig:policy-example}(a), ``this information'' would be normalized to ``geolocation'', same as the phrase ``location'' that it refers to.
Please see~\Cref{appendix:phrase-normalization} for details on the phrase normalization.

\paragraph{Unspecified Data and Unspecified Third Party}
As a special case of phrase normalization, \privacypolicy{s} often use blanket terms like ``information'' and ``third party'' without further details.
For example, in the sentence ``We collect information to provide services'', the word ``information'' can be interpreted as unspecified (or all possible) data types.
We find it more appropriate to treat such blanket terms specially than assuming them to have consistent meaning across the text.
\tool{} uses two special nodes, ``unspecified data'' for data types and ``unspecified third party'' for entities, as the normalized forms of such blanket terms in \poligraph{}s.

\paragraph{Classifying Purposes}
The purpose annotator identifies purpose phrases. To allow automated analysis of purpose, we coarsely group purpose phrases into five categories:
\textit{services},
\textit{security},
\textit{legal},
\textit{advertising}, and
\textit{analytics}.
These categories are derived from the business and commercial purposes defined in the CCPA~\cite{ccpa}. %
As in prior work~\cite{trimananda2022ovrseen}, we distinguish between \textit{core} (\ie{} services, security, and legal) and \textit{non-core} (\ie{} advertising and analytics) purposes of data collection.

We fine-tune a sentence transformer model~\cite{reimers-2019-sentence-bert} to classify purpose phrases into these categories. For example, the phrase ``to provide features'' is classified as \textit{services}, whereas the phrase ``for advertising purposes'' is classified as \textit{advertising}.
Note that a purpose phrase can be classified into multiple labels if it mentions more than one purpose.
To train the model, we manually annotate a dataset of 200 phrases. We use SetFit~\cite{tunstall2022efficient} to enable few-shot fine-tuning with this small dataset.
This can be done because the underlying transformer model is already trained on a large corpus to gain language knowledge, and SetFit takes advantage of contrastive learning to learn the differences between classes effectively.
The performance of purpose classification is reported in \Cref{sec:evaluation}.

\paragraph{Building \poligraph}
Finally, \tool{} builds the \poligraph{} from the phrase graph by merging phrases with the same normalized form into one node, keeping \COLLECT{} and \SUBSUME{} edges, and inferring the $Purposes(\cdot)$ attributes 
from \PURPOSE{} edges in the phrase graph.
Because NLP errors occasionally cause invalid edges, such as circles, \tool{} validates nodes and edges to ensure that the final graph conforms with the Definition~\ref{def:poligraph} of \poligraph{}.

\par
\begin{figure*}[!t]
    \begin{minipage}{0.50\textwidth}
        \centering
        \includegraphics[width=\textwidth,page=3]{figures/poligraph-figures-crop.pdf}
        \caption{The \poligraph{} generated from `Puzzle 100 Doors'' app's \privacypolicy{}.}
        \label{fig:full-poligraph}
    \end{minipage}\hfill
    \begin{minipage}{0.46\textwidth}
        \centering
        \includegraphics[width=\textwidth,page=1]{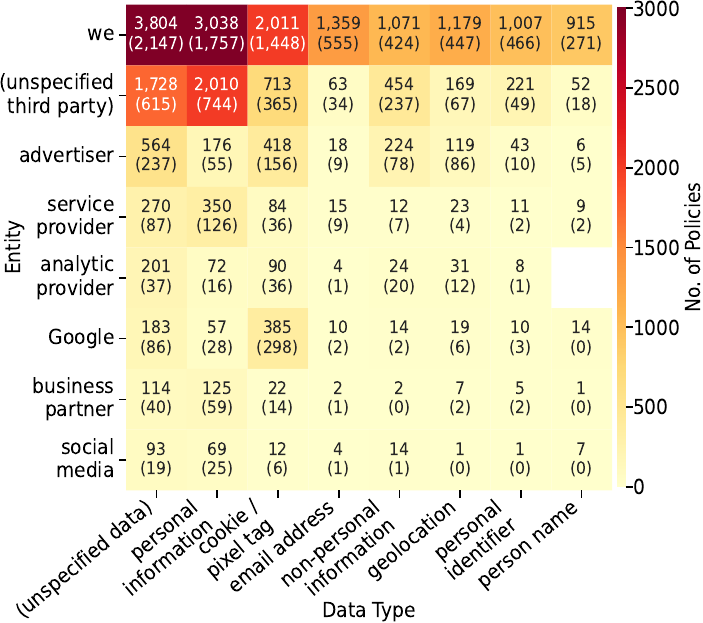}
        \caption{Statistics of common \COLLECT{} edges. The numbers of edges that have $Purposes(\cdot)$ attributes are shown in parentheses.}
        \label{fig:common-collect-edges}
        \includegraphics[width=\textwidth,page=2]{figures/result-policy-crop.pdf}
        \caption{Statistics of common \SUBSUME{} edges between data types.}
        \label{fig:common-subsume-edges}
    \end{minipage}
\end{figure*}

\vspace{\baselineskip}
\Cref{fig:full-poligraph} shows an example of a \poligraph{} generated from a simple policy~\cite{puzzle100doors-policy} for demonstration purposes.
A typical \poligraph{} from the policies that we have analyzed (see \Cref{sec:applications}) can contain up to hundreds of nodes and edges.
It is common to see vague phrases like ``statistical user data'' that are not further clarified, and misleading definitions like claiming anonymized information to include data types that are likely personal and identifiable. However, it is important that \poligraph{} does capture data collection, along with its purposes, and subsumptions for further analysis.

\subsection{Evaluation}
\label{sec:evaluation}

In this section, we evaluate \tool{}'s performance in analyzing policies. At the beginning of each subsection, we state the research question (RQ) addressed therein, our approach, and a preview of the results.

\paragraph{The PoliCheck Dataset} Throughout this section and \Cref{sec:applications,sec:llmtool}, we use the public dataset provided by PoliCheck~\cite{andow2020actions}. %
We choose this dataset because it is among the largest public datasets for policies. The dataset consists of policies of 13,796 Android apps on Google Play Store. The number of policies is large enough to necessitate automated analysis.
Furthermore, it also comes with the apps' network traffic data that facilitate flow-to-policy consistency analysis in Section~\ref{sec:flow-to-policy-consistency-analysis}. We write a crawler script based on the Playwright library~\cite{playwright} to download the policy text from each URL. We obtain the most recent version of the policies from March 2023. After excluding non-English, invalid, and duplicate webpages, we obtain \result{6,084} unique policies used by \result{13,626} apps.

\paragraph{Test Set} Out of the full PoliCheck dataset, we randomly select a set of 200 policies, and we read and annotate them in order to build ground truth for various evaluation tasks. This test set has no overlap with the data used to generate the NER corpus or to train the purpose classifier (see Section~\ref{sec:implementation}).

\subsubsection{\poligraph{} Generation and Characterization}
\label{subsec:poligraph-generation}

\tool{} successfully generates \poligraph{}s for \result{5,255} policies. The remaining policies cannot be processed because they either do not claim to collect data, or use irregular or unsupported HTML tags that cannot be correctly parsed.
We characterize the \COLLECT{} and \SUBSUME{} edges in the \poligraph{}s generated by \tool{} as follows.

\paragraph{\COLLECT{} edges} \tool{} extracts \result{103,185} \COLLECT{} edges from \result{100,565} sentences in total. Among them, \result{34,052} edges have $Purposes(\cdot)$ attributes from \result{38,994} purpose phrases.
Figure~\ref{fig:common-collect-edges} shows the common \COLLECT{} edges found in the dataset.
Generic terms, such as ``personal information'' and ``personal identifier'', are commonly used to express data types in policies. %
Some specific terms, such as ``cookie / pixel tag'', ``email address'', and ``ip address'' are also found in many policies.
Furthermore, we find that policies disclose data collection  by first-party (\ie{} ``we'') more frequently than by third-party entities. Major third-party entity categories are ``advertiser'' and ``analytic provider''. Google, as the platform, is also frequently mentioned in the policies.

\paragraph{\SUBSUME{} edges}  \tool{} extracts \result{52,007} \SUBSUME{} edges from \result{20,959} sentences in the dataset. Figure \ref{fig:common-subsume-edges} shows common \SUBSUME{} edges that connect data type nodes. ``Personal information'', ``contact information'' and ``personal identifier'' are the most frequently used generic terms to represent data types. Notably, we find that many policies declare the collected data as ``non-personal information'': this conflicts with our CCPA-based global data ontology. We will discuss the issue of misleading definitions in Section~\ref{sec:terms-and-definitions}.

\paragraph{``Unspecified''} The nodes ``unspecified data'' and ``unspecified third party'' (see Section~\ref{sec:building-knowledge-graph}) are found in \result{72.0\% (3,785)} of \poligraph{}s. This is because many policies discuss data collection, sharing, and use in separate sections. When they discuss sharing, precise data types are often omitted. When they discuss purposes of use, both data types and entities can be unspecified terms. For example, KAYAK's policy states: ``To protect rights and property, we may disclose your information to third parties''~\cite{kayak-policy}. Without further details on ``information'' and ``third parties'', the statement is captured in the \poligraph{} as \textit{unspecified third party}\edge{COLLECT}\textit{unspecified data} with \textit{security} as the purpose.

\subsubsection{Manual Validation of \poligraph{}s}
\label{sec:edge-validation}
How accurate is \tool{} in generating \poligraph{}s from policies \wrt{} the definitions in Section~\ref{sec:terminology}?
To answer this question, we manually evaluate whether \tool{} extracts the correct edges. To evaluate the \textit{precision} of \poligraph{} edges, we sample five edges from each of 100 randomly selected \poligraph{}s in the dataset and read the corresponding policy text to validate whether each edge is correctly extracted from the text. To help with this evaluation, \tool{} stores the sentences from which each edge is generated. 
We find that the precision for \COLLECT{} edges is \result{90.4\%}, and the precision for \SUBSUME{} edges is \result{87.7\%}.
We do not evaluate recall as it turned out to be difficult for humans to label edges without biases for all data types and entities. We will report recalls in Section~\ref{subsec:comparison-with-policylint} where we consider a subset of common data types.

In theory, false positive edges can propagate to more incorrect inferences. For example, a false \textit{advertiser}\edge{COLLECT} \textit{personal information} edge would lead to wrong inferences of \textit{advertiser} collecting all data types subsumed by \textit{personal information}. However, we find that such cases are rare, and false positives are often caused by recognizing irrelevant phrases as data types or entities.
For example, in the sentence ``the app may use third party code'', \tool{} mistakes ``third party code'' as a data type.
This is less of an issue if we scope the analysis to a subset of common data types or entities (as in Section~\ref{subsec:comparison-with-policylint}). Most false-positive edges come from NLP errors including: (1) NER recognizing irrelevant phrases as data types or entities, and (2) mistaking some interrogative or negative sentences as affirmative statements. 

We also evaluate \tool{}'s purpose classification model.
We randomly sample five purposes phrases identified by the purpose classifier from each policy in the test set, manually assign labels to the phrases, and compare them to the ones labelled by the purpose classification model.
Overall, the \textit{macro-averaged precision} and \textit{recall} are \result{91.0\%} and \result{94.8\%}, respectively, for this multi-label multi-class classification task.

\subsubsection{Comparison to Prior Policy Analyzers}
\label{subsec:comparison-with-policylint}
How well does \tool{} analyze collection statements compared to prior state-of-the-art policy analyzers?
To answer this question, we use \poligraph{}s to infer collection relations, namely $collect(n, d)$ indicating that an entity $n$ may collect a data type $d$.
We obtain tuples of collection statements from prior state-of-the-art, namely PolicyLint~\cite{andow2019policylint}. We compare both results to manually labeled ground truth.

\paragraph{Methodology}
Since PolicyLint extracts tuples (see Figure~\ref{fig:policy-example}(b)), we convert pairs of $collect(n, d)$ relations in \poligraph{}s into PolicyLint tuples $\langle$\textit{n, collect, d}$\rangle$. Still, we cannot compare data types and entities from the two tools directly because they normalize phrases in different ways. To work around the issue, we select a subset of terms to compare. For data types, we only consider the following precise data types in PolicyLint, since they are comparable to the same data types extracted by \tool{}:
``mac address'', ``router ssid'', ``android id'', ``gsf id'',
``sim serial number'', ``serial number'', ``imei'', ``advertising identifier'', 
``email address'', ``phone number'', ``person name'', and ``geographical location''.\footnote{Note that we map ``coarse geolocation'', ``precise geolocation'', and ``geolocation'' in \poligraph{} all to ``geographical location'' in the tuple because PolicyLint does not distinguish between them.} 
For entities, we only distinguish between the first party and third party, \ie{} all tuples are converted to either $\langle$\textit{we, collect, data type}$\rangle$ or $\langle$\textit{third party, collect, data type}$\rangle$; ``unspecified third party'' in \poligraph{} is considered a third party.

\tool{} finds \result{13,529} tuples in the entire dataset. PolicyLint finds \result{6,410} tuples.
\revision{To evaluate the precision and recall, we manually verify all the tuples (collection relations) extracted by \tool{}, PolicyLint, \llmtool{} (see~\Cref{sec:llmtool}), and we further use coarse regular expressions to match all possible mentions of the 12 data types in policies with a likely high chance of false positives.
Two of our authors read the text to determine if each data type is collected by the first party or any third party to create the ground truth\footnote{To save time, we did not read the entire privacy policy text. Therefore, it is possible that we omit some mentions of data types and overestimate the recall. Also, due to the inclusion of new results from~\Cref{sec:llmtool}, the recall scores reported here are different from~\citet{usenix-version}}.}

\paragraph{Precision} \tool{} achieves 97.2\% precision, and PolicyLint achieves 93.1\% precision. As previously explained, NLP errors are the main reason of wrong collection relations.
We improve the precision by using recent NLP models.
The precision of \tool{} is higher than reported in Section~\ref{sec:edge-validation} because here we only consider a subset of data types and thus many falsely labeled data types are excluded.
Also note that both tools show lower precision for third-party tuples (see Table~\ref{tab:ablation-study}) because some policies use company names rather than ``we'' to refer to the first party, and both tools can mistake the company names as third parties in this case.

\revision{
Although \tool{} uses state-of-the-art linguistic analysis models, there are still NLP errors for various reasons.
First, NLP models have difficulties in parsing long and complicated sentences, which are common in privacy policy documents.
As discussed in \Cref{sec:edge-validation}, the NER model can classify %
irrelevant phrases as data types or entities, resulting in meaningless \COLLECT{} or \SUBSUME{} edges like \textit{we}\edge{COLLECT}\textit{third-party code}.
Furthermore, interrogative or negative sentences are occasionally mistaken as affirmative ones, which result in wrong edges.
Second, due to the graph structure of \poligraph{}, false-positive edges can propagate to more incorrect inferences of subsumption and collection relations as discussed in \Cref{sec:edge-validation}.
Most of our analysis in \Cref{sec:evaluation,sec:applications} focuses on a limited set of well-known data types and/or entities, so irrelevant terms are naturally excluded.
Given the relatively high precision, the issue of misinterpreting some policies should not significantly impact the validity of overall trends discussed in \Cref{sec:applications}.
}

\paragraph{Recall} \tool{} achieves 65.5\% recall, a significant improvement from PolicyLint's 27.2\% recall.
As we will explain later in~\Cref{subsec:ablation-study}, the graph structure and improved NLP techniques both contribute to the higher recall.

\revision{
Despite the improvement, the recall of our tool is still limited, mainly by its linguistic analysis approach.
First, \tool{} relies on linguistic analysis, namely analyzing syntactic features of sentences, to extract relations. As a result, a list 
such as ``Information we collect: • Name; • Gender'',  which contains no complete sentence, cannot be processed. \tool{} cannot handle tables either, despite the increasing use of tables in policies for readability.
This limits the recall of our tool in discovering relations.
Second, \tool{}'s annotators rely on manually written rules of syntactic patterns to extract data types, entities and purposes in the text. The flexibility of human language makes it difficult to cover all possible patterns.
For example, the collection annotator does not recognize the sentence
``Among the types of Personal Data that we collect..., there are: Cookies...''\cite{wordsnack-privacy-policy}, because there is no rule to cover the odd statement. While it is possible to write a new rule for it, we decided to only include common and intuitive patterns (see \Cref{sec:annotators}) in our implementation, for simplicity.
}

\par\smallskip
Given the high precision and improved, but imperfect, recall, we recommend to interpret what \tool{} captures as a lower bound of the actual collection statements.

\begin{table}[t!]
\centering
\caption{\revision{Manual validation and ablation studies results}.}
\label{tab:ablation-study}
\centering
\small{
\begin{tabular}{@{}lll@{\hspace{1em}}l@{}}
\toprule
\multicolumn{2}{r}{\textbf{\# tuples}} & \textbf{precision (1st/3rd party)} & \textbf{recall (1st/3rd party)} \\ \midrule
\textbf{Manual Validation}        &     &                          &                          \\
Ground Truth                        &   950 & -                        & -                        \\
\tool{}                           &   640 & 97.2\% (99.8\% / 92.8\%) & 65.5\% (62.8\% / 71.1\%) \\
PolicyLint                        & 291 & 93.1\% (94.2\% / 85.3\%) & 27.2\% (37.7\% / 8.2\%) \\
\textbf{Ablation Studies}           &     &                          &                          \\
\textit{no-subsumption-annotator}    & 345 & 97.4\% (100\% / 91.3\%) & 33.7\% (37.5\% / 26.8\%) \\
\textit{no-coreference-annotator}    & 616 & 97.1\% (99.7\% / 92.3\%)  & 60.0\% (61.2\% / 57.9\%) \\
\textit{no-list-annotator}        & 614 & 97.2\% (99.7\% / 92.9\%)  & 59.9\% (60.3\% / 59.3\%) \\
\textit{per-sentence-extraction}  & 471 & 97.0\% (99.7\% / 90.5\%)  & 45.9\% (51.9\% / 35.0\%) \\
\textit{per-section-extraction}   & 573 & 97.4\% (99.7\% / 92.6\%)  & 56.0\% (59.8\% / 49.2\%) \\ \bottomrule
\end{tabular}
}

\end{table}

\subsubsection{Ablation Studies}
\label{subsec:ablation-study}
\poligraph{} consists of many components. Where does the performance improvement come from?
To answer this question, we conduct ablation studies to understand how each component and design decision contributes to \tool{}'s performance improvements. We modify \poligraph{} into the following experimental configurations, and we summarize the evaluation results in Table~\ref{tab:ablation-study}.

\paragraph{Removing Components} In the \textit{no-subsumption-annotator}, \textit{no-coreference-annotator}, \textit{no-list-annotator} configurations, we disable one component at a time and assess the effect.
As shown in Table~\ref{tab:ablation-study}, 
disabling the subsumption annotator reduces recall from 65.5\% to 33.7\%. The reason is that the precise data types that we evaluate are often subsumed by generic terms. They have to be linked by \SUBSUME{} edges to allow inferences of collection relations. Each of the coreference and list annotators contributes about 5\% to recall.

\paragraph{Limiting Extraction Area} In the \textit{per-sentence-extraction} configuration, we modify \tool{} to behave like PolicyLint and extract tuples within sentence boundaries\footnote{The \textit{per-sentence-extraction} is different from removing \SUBSUME{} edges because PolicyLint ideally is still able to find data types and entities that are subsumed by generic terms within the same sentence.}.
This is done by filtering out $collect(n, d)$ relations in \poligraph{}s where the entity and data type comes from different sentences.
Similarly, \textit{per-section-extraction} only considers collection relations within the boundaries of sections.
As explained in Section~\ref{sec:nlp-on-structured-document}, \tool{} keeps headings from HTML, which is considered as the boundaries of sections here.
These configurations help assess the improvement by introducing the graph structure to infer relations across sentences and sections.

Table~\ref{tab:ablation-study} shows the results. The \textit{per-sentence-extraction} configuration has only 45.9\% recall compared to \tool{}'s 65.5\%. On one hand, the sophisticated NLP methodology still improves performance over PolicyLint. On the other hand, the graph structure is necessary to infer 19.6\% of all relations. The graph is more effective in identifying third-party data collection---it allows us to find 36.1\% of third-party tuples, because the disclosure of third-party sharing often uses broader terms (\eg{} ``\textit{Anonymous information} may be shared with analytic providers'') and the exact data types (or entities) must be inferred through \poligraph{}.
The \textit{per-section-extraction} configuration, by connecting data types and entities within a longer textual region, achieves 56.0\% recall but still falls behind the full version by missing 9.5\% of relations. As explained in Section~\ref{sec:introduction}, the typical structure of policies that discuss data collection, use and sharing in separate sections makes it necessary to find exact data types and entities in different sections.
Therefore, we argue that \poligraph{} enables a much better coverage by connecting information disclosed in different sentences and sections to infer more collection relations in a policy.%

\section{\poligraph{} Applications}
\label{sec:applications}

In this section, we present two novel applications enabled by \poligraph{}. Section~\ref{sec:summarization} presents policies summarization, which provides inferences on the common patterns across different policies. Section~\ref{sec:terms-and-definitions} looks into how the same or similar terms are defined across different policies.
In addition, we show that \poligraph{} can improve two applications that have been explored by prior work. Section~\ref{sec:negative-statements} extends \poligraph{} to identify contradicted statements. Section~\ref{sec:flow-to-policy-consistency-analysis} applies \poligraph{} to check the consistency between the data flows of a mobile app and its policy.

\begin{figure*}
     \centering
     \begin{subfigure}[b]{0.32\textwidth}
         \centering
         \includegraphics[width=\columnwidth,page=3]{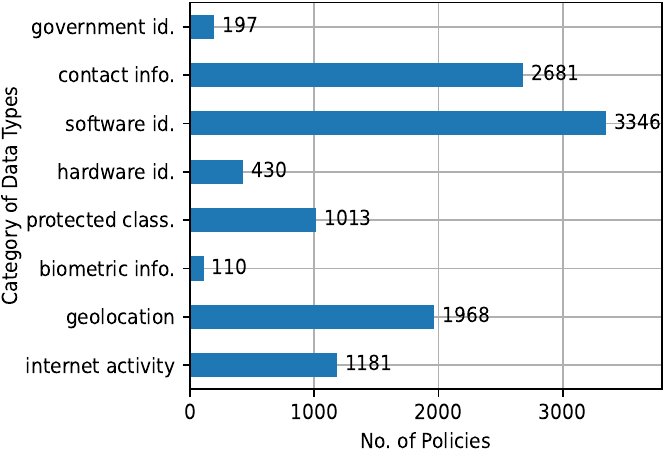}
         \captionsetup{skip=2pt}
         \caption{Collection of Data Types}
         \label{fig:stat-collection}
     \end{subfigure}
     \hfill
     \begin{subfigure}[b]{0.33\textwidth}
         \centering
         \includegraphics[width=0.95\columnwidth,page=4]{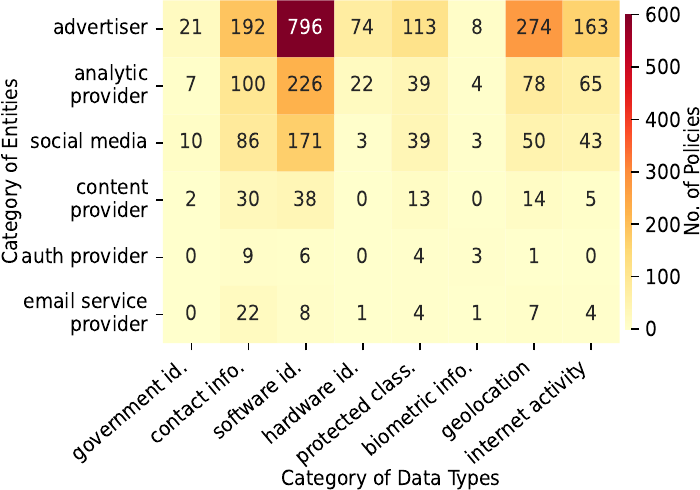}
         \captionsetup{skip=2pt}
         \caption{Third-party Entities of Data Collection}
         \label{fig:data-sharing-heatmap}
     \end{subfigure}
     \hfill
     \begin{subfigure}[b]{0.34\textwidth}
         \centering
         \includegraphics[width=\columnwidth,page=5]{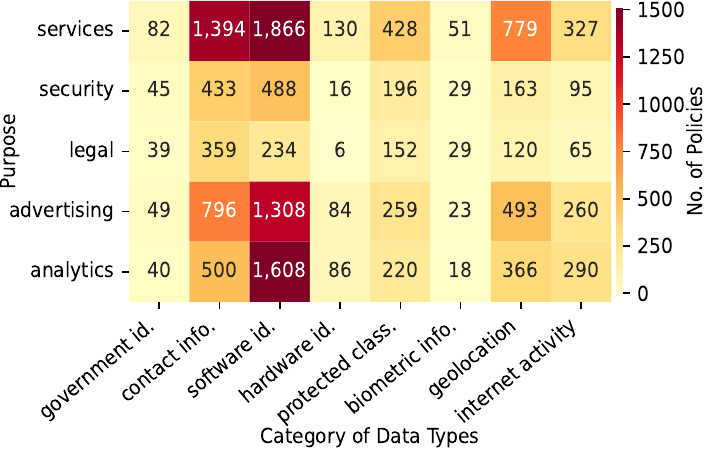}
         \captionsetup{skip=2pt}
         \caption{Purposes of Data Collection}
         \label{fig:data-purpose-heatmap}
     \end{subfigure}
     \caption{Policy Summarization --- Statistics of policies that disclose
     \textbf{(a)} the collection of eight categories of data types;
     \textbf{(b)} data collection by third-party entities (per data type category); and
     \textbf{(c)} the purposes for the collection (per data type category).
     For example, 3,348 policies claim to collect data types under the ``software identifier'' category;
     794 policies disclose that data types in this category are collected by ``advertiser'' as the third-party entity.
     And 1,847 policies disclose the collection of this category for the ``services'' purpose.
     }
     \label{fig:policy-summarization-results}
\end{figure*}

\subsection{Policies Summarization}
\label{sec:summarization}

We use \poligraph{} to summarize {\em all} policies in our dataset and reveal common patterns among them. Specifically, we aim to identify: \textit{(1) how common each category of data types is collected;}
\textit{(2) what kind of entities collect these data types;}
and \textit{(3) the purposes for which these data types are used.}
As data types and entities captured by \poligraph{}s are unorganized and differ across policies (see Section~\ref{sec:edge-validation} for examples), we use our global ontologies to categorize data types and entities in a canonical manner.

\paragraph{Data Types}
In this analysis, we use the eight parent nodes of the leaf nodes in the data ontology shown in  Figure~\ref{fig:data-ontology} to group the data types into eight categories%
\footnote{\label{ftnote:unknown-terms}``Unspecified data'', ``unspecified third party'' and other data types and entities that are not part of the global ontologies (see Section~\ref{sec:ontology}) are excluded from the analysis of policies summarization.}:
``government identifier'', 
``contact information'', 
``software identifier'', 
``hardware identifier'', 
``protected classification'', 
``biometric information'',
``geolocation'', and 
``internet activity''.
Figure \ref{fig:stat-collection} shows the numbers of policies that disclose the collection of data types in each category.
Overall, \result{77.9\% (4,093)} of policies disclose the collection of at least one of these data categories.

\begin{finding}{}
\label{finding:device-id-distribution}
\textit{The most frequently collected data category is ``software identifier'', which mostly originates from ``cookie'' as the specific data type being collected.}
\result{63.7\% (3,346)} of policies disclose the collection of ``software identifier''.
Among the specific data types, ``cookie / pixel tag'' is the most common and found in \result{80.3\% (2,688 / 3,346)} of these policies. On the other hand, identifiers specific to mobile apps, mainly ``advertising ID'' and ``Android ID'', are found in only \result{26.0\% (870) and 2.8\% (95)} of these policies, respectively.
Many developers simply write one policy for various products, including mobile apps and web-based services. Furthermore, some developers seem to use ``cookie'' as a generic term for all kinds of device identifiers for tracking.
\end{finding}

\paragraph{Third-Party Entities}
We use the six parent nodes of the leaf nodes in the entity ontology shown in Figure~\ref{fig:entity-ontology} to group entities into six categories\footref{ftnote:unknown-terms}: 
``advertiser'', 
``analytic provider'', 
``social media'',
``content provider'',
``auth provider'', and
``email service provider''.
Figure~\ref{fig:data-sharing-heatmap} reports how each data type category is disclosed to be shared with or collected by these third-party entities (\ie{} $\mathit{collect}(n, d)$ relations).

\begin{finding}{}
\label{finding:sharing-usage-difference}
\textit{``Software identifier'' is frequently shared with advertisers. Third-party sharing of other data categories (\eg{} ``geolocation'', ``protected classification'',
and ``internet activity'') is also non-negligible.}
We find that \result{23.8\% (796 / 3,348)} of policies that disclose to collect ``software identifier'' involve sharing with advertisers. Analytic providers and social media are other major third parties with whom apps share data.
\result{620} policies share data in other categories (\eg{} ``geolocation'', ``internet activity'', and even ``protected classification'') with third parties.
As data in these categories may be sensitive, it is doubtful whether the sharing of them is appropriate.
\end{finding}

\begin{finding}{}
\label{generic-term}
\textit{Many policies disclose data sharing using generic terms, \eg{} ``personal information''. This leads to the inference that the app may share all data types that a generic term subsumes.}
This often happens when policies disclose data collection, sharing, and use separately. For example, Figure~\ref{fig:policy-example} shows that ``personal information'' is shared with entities such as ``social networking services''. This may be alarming to users since ``personal information'' subsumes sensitive data types, such as ``location'' and ``IP address''.
The use of generic terms reduces transparency. Users are left wondering which, if not all, ``personal information'' is shared.
We find that 710 policies declare third-party collection or sharing using generic terms that subsume data types in multiple categories.
\end{finding}

\paragraph{Purposes} Figure~\ref{fig:data-purpose-heatmap} reports the statistics of policies that disclose purposes of data collection, as discussed in Section~\ref{sec:building-knowledge-graph}, per data type category. 

\begin{finding}{}
\label{finding:non-core-usage}
\textit{\result{56.9\% (2,990)} policies disclose that their apps collect data for non-core purposes. In over 80\% of them \result{(2,398)}, the main non-core purpose is advertising.}
We find that, while ``software identifier'' remains the most common data type category used for non-core purposes, the potential use of other data types for non-core purposes is concerning. For instance, the collection of ``geolocation'', ``protected classification'', and ``internet activity'' for non-core purposes is declared in about 10\% of policies.
The CCPA~\cite{ccpa} defines ``government identifiers'', ``precise geolocation'', and certain protected classifications as \textit{sensitive personal information}---the law limits the usage of these data (\eg{} users have the right to limit the use of such personal information for non-core purposes).
\end{finding}

\subsection{Correct Definitions of Terms}
\label{sec:terms-and-definitions}

The second novel application enabled by \poligraph{} is assessing the correctness of definitions of terms. Besides summarizing policies on their own right,
we can check whether a policy defines terms in ways that are consistent with external knowledge as captured by global ontologies. %
This is necessary because policies often provide their own definitions of terms. This is not a problem if the definitions align with external knowledge (\eg{} privacy laws), but it may be \textit{misleading} if they do not agree. For example, some policies define ``geolocation'' as ``non-personal information''. 
In this section, we check whether the definitions of data type terms in individual policies (as captured by their \poligraph{}s' local ontologies) align with our CCPA-based global data ontology (see Figure \ref{fig:data-ontology}).

Overall, such different definitions are found in \result{25.5\% (1,339 / 5,255)} policies in our dataset, as listed in Table~\ref{tab:different-definitions}.

\begin{table}[t!]
\centering
\caption{Examples of different definitions found in \poligraph{}s with respect to the global data ontology. For example, ``geolocation'' is defined as ``non-personal information'' in 123 policies. 
}
\label{tab:different-definitions}
\centering

\small{
\begin{tabular}{@{}p{32mm} p{90mm}@{}}
\toprule
\textbf{Hypernym} &
  \textbf{Hyponym (\# Policies)} \\ \midrule
non-personal information & 
{
ip address            (126),
geolocation           (123),
device identifier     (108),
gender                (76),
application installed (72),
age                   (70),
identifier            (46),
internet activity     (44),
device information    (38),
coarse geolocation    (35) ...
} \\ \midrule

aggregate / deidentified / pseudonymized information &
{
ip address                (122),
device identifier         (89),
geolocation               (78),
browsing / search history (16) ...} \\ \midrule

internet activity
& {
ip address         (151),
device identifier  (107),
geolocation        (40),
advertising id     (13),
cookie / pixel tag (10) ...}\\ \midrule

geolocation &
{
ip address       (76),
postal address   (15),
router ssid      (10) ...} \\ \midrule

personal identifier &
{
advertising id      (74),
cookie / pixel tag  (49),
device identifier   (39),
geolocation         (35),
date of birth       (27),
gender              (23) ...} \\ \bottomrule

\end{tabular}
}

\end{table}

\begin{finding}{}
\textit{Many policies define data types that they collect to  be ``non-personal'', ``aggregated'', ``deidentified'', or ``pseudonymized''. However, this can be inconsistent with the definitions in the CCPA.}
Indeed, in the CCPA, ``deidentified information'' is defined as information that ``cannot reasonably identify, relate to, describe ...
to a particular consumer''.
Although entities technically can deidentify personal information, some of the data types we observe in Table~\ref{tab:different-definitions}, notably ``geolocation'', ``gender'', ``age'', and ``date of birth'', are generally considered personal information by the public and according to the CCPA.
Declaring these data types as \textit{non-personal} or \textit{deidentified} can be misleading. %
For example, Paleblue declares in its policy~\cite{paleblue-privacy-policy} that ``Paleblue may also invite you to share \textit{non-personal information} about yourself which may include... (1) your age or date of birth; (2) your gender...''.
\end{finding}

\begin{finding}{}
\textit{Many policies use non-standard terms. They can have broad or varied definitions across different policies.}
For example, it is not surprising that the definition of ``profile information'' is application-specific. One policy from the Manager Readme app~\cite{manager-readme-privacy-policy} 
defines ``profile information'' to include ``name'' and ``location'', while another policy from Armor Game Inc.~\cite{armor-game-privacy-policy}
defines the term to include ``gender'' and ``birthday''. In these cases, the use of non-standard terms is acceptable as the policies clearly explain what they mean by the terms. However, we also find many policies that do not clearly define their non-standard terms. Particularly, while ``profile information'' is found in 178 policies, subsumption relationships are found in only 17 of them in their corresponding \poligraph{}s.
Table~\ref{tab:non-standard-terms} presents examples of non-standard terms and their possible definitions found in the policies.%
\end{finding}

\begin{table}[t!]
\centering
\caption{Examples of non-standard terms found in policies. For example, ``technical information'' is used in 311 policies but its detailed definition is only found in 126 policies. }
\label{tab:non-standard-terms}
\centering
\small{
\setlength\tabcolsep{3pt}
\begin{tabular}{@{}p{38mm} p{90mm}@{}}
\toprule
\textbf{Term {(\# Policies)}} &
  \textbf{Possible definitions found in policies} \\ \midrule
technical information (311) &
  {\textit{From 126 policies:}
  advertising id, age, android id, browsing / search history, cookie / pixel tag, device identifier, email address, geolocation, imei, ip address, mac address ...}
  \\ \midrule
profile information (178) &
  {\textit{From 17 policies:}
age, contact information, date of birth, email address, gender, geolocation, person name, phone number ...}
  \\ \midrule
demographic information (315) &
  {\textit{From 112 policies:}
age, browsing / search history, date of birth, email address, gender, geolocation, ip address, postal address, precise geolocation, race / ethnicity, router ssid ...}
  \\  \midrule
log data (81) &
  {\textit{From 52 policies:}
advertising id, android id, cookie / pixel tag, coarse geolocation, cookie / pixel tag, email address, geolocation, imei, ip address, mac address, person name ...}
  \\  \bottomrule
\end{tabular}
}

\end{table}

\subsection{Contradiction Analysis} 
\label{sec:negative-statements}

In this section, we apply \tool{} to analyze contradictions within a policy.
To that end, we extend \poligraph{} to also analyze negative collection statements (\eg ``We do not collect personal data'') ignored in previous sections (which only analyzed affirmative statements). We also propose extensions to the main \poligraph{} framework so as to capture additional contexts that are crucial to interpret contradictions.

Prior work, namely PolicyLint~\cite{andow2019policylint}, identifies contradictions by detecting affirmative and negative sentences that mention the same or conflicting entities and data types.
By manually checking all the 86 contradictions identified by PolicyLint in our test set, we found 79 of them turn out to be false alarms.
This is because the analysis ignores many contexts surrounding the ``contradicted'' statements, as explained below.

\begin{compactitem}[$\bullet$]
\item{\em Different Purposes:} 
PolicyLint considers ``we collect PII to provide the service'' and ``we do not collect PII for advertising purposes'' as contradicting statements, despite them discussing data collection for different purposes, \ie{}  services \vs{} advertising.

\item{\em Different Data Subjects:} 
PolicyLint considers ``we collect PII'' and ``we do not collect PII from minors'' as contradicting. A human reader would recognize that the sentences  discuss different data subjects, \ie{} general user \vs{} child.%
\footnote{Beyond semantics, another source of error for PolicyLint was its implementation. In order to find and skip text about children policy, which was considered outside the framework, it performed string matching using hardcoded regular expressions, which often failed due to sentence variability.}

\item{\em Different Actions:} 
PolicyLint considers ``... share PII with third parties'' and ``... do not sell PII to any third party'' as contradictions. A human reader would recognize that they refer to different actions, \ie{}  share \vs{} sell.

\item{\em Contradictions According to Global Ontologies:}
In addition, some policies do not literally contradict themselves, but the data types and entities in the affirmative and negative statements overlap according to PolicyLint's ontologies.
For example, in the policy of Horizone Media~\cite{horizone-media-privacy-policy}, PolicyLint reports ``we use anonymous identifiers'' and ``we do not collect personal information'' as a contradiction. The policy does not define ``anonymous identifiers'' as ``personal information'', but PolicyLint views all ``identifiers'' as ``personal information'' in its data ontology. In this case, the policy can be considered as misleading but does not directly contradict itself. Note that we already discuss misleading definitions in Section~\ref{sec:terms-and-definitions}.
\end{compactitem}

\subsubsection{Framework Extensions}
\label{subsubsec:negative-statements-extensions}

The main \poligraph{} framework, described in Section \ref{sec:knowledge-graph}, introduced data types, entities and purposes.
However, it only deals with affirmative, not negative, statements. In this section, we extend \poligraph{} to also analyze negative statements, and we show how to deal with contradicted statements in fine-grained contexts.
To that end, we add a new type of negative edge (\NOTCOLLECT{}), actions as edge subtypes, and the notion of data subject, as described next.%
\footnote{The extensions are presented in this section, as opposed to as part of the main framework, for several reasons. First, they are specific to contradiction analysis, and motivated by the limitations of prior work in taking into account fine-grained contexts in that analysis.
Second, as shown in Section~\ref{subsubsec:contradiciton-results}, although effective in refining contexts for contradiction analysis, these extensions do not address {\em all} aspects of fine-grained contexts: \eg{} one can define additional types of actions, data subjects, and other contexts.
Third, these extensions do not affect the validity of previous results in Sections~\ref{sec:evaluation} and~\ref{sec:applications}, which were based only on affirmative, and ignored negative, statements.}

\paragraph{Negative Collection Statements}

The collection annotator (see Section~\ref{sec:implementation}) identifies negative sentences and by default excludes them. In this section, we modify the annotator to also account for negative collection statements and represent them as \NOTCOLLECT{} edges.
For example, ``we do not collect personal information'' will be represented with edge \textit{we}\allowbreak\edge{NOT\_COLLECT}\allowbreak\textit{personal information} in \poligraph{}. Similarly to its positive counterpart, a \NOTCOLLECT{} edge can have $Purposes(\cdot)$ attributes.

\paragraph{Refining Actions}
We further consider five subtypes of  \COLLECT{} (and \NOTCOLLECT{}) edges that represent different data actions: $\{$\textit{collect}, \textit{be\_shared}, \textit{be\_sold}, \textit{use}, \textit{store}$\}$.
We denote an action-sensitive \COLLECT{} edge as $n$\edge{COLLECT $[a_{pos}]$}$d$, and an action-sensitive \NOTCOLLECT{} edge as $n$\edge{NOT\_COLLECT $[a_{neg}]$}$d$, where $a_{pos}$ and $a_{neg}$ are one of the 5 subtypes of data actions.

We extend the collection annotator to map verbs to these subtypes accordingly.
For example, ``... do not sell personal information to advertisers'' is represented as edge \textit{advertisers}\edge{NOT\_COLLECT $[be\_sold]$}\textit{personal information} in \poligraph{}. 
Table~\ref{tab:verb-to-actions} in~\Cref{appendix:contradiction-framework-extension} provides the list of verbs and corresponding actions, which can be extended if so desired.

\paragraph{Data Subjects}
We extend data type nodes to include subjects, \ie{} the group of people to which the data type pertains, following the terminology of Contextual Integrity~\cite{nissenbaum2009privacy}. We denote a subject-sensitive data type node as a pair $(d, s)$, where $d$ is the data type, and $s$ is the subject of the collected data.

We add a new subject annotator in \tool{} to identify subjects of data types. Currently, we implement it to identify a commonly-seen data subject: children. In the current implementation, we define subject $s \in \{\textit{child, general user}\}$, and we represent the statement ``we don't collect personal information from children'' as  \textit{we}\edge{NOT\_COLLECT $[collect]$}(\textit{personal information, child}) in \poligraph{}. We note, however, that the modular implementation of \tool{} allows for the set of subjects to be extended to include additional subjects.

\subsubsection{Contradiction Analysis}
\label{subsubsec:negative-statements-analysis}

To identify potential contradictions, we assess whether a positive edge and a negative edge in a \poligraph{} involve conflicting %
 data types, subjects, entities, actions, and purposes.
A pair of edges may contradict if all these parameters conflict; otherwise, the two statements address different aspects of data collection, as outlined earlier, thus do not contradict.

\begin{definition}{\textbf{\textit{Conflicting Edges.}}}
A positive edge $e_{pos} = n_{pos}$\edge{COLLECT $[a_{pos}]$}$(d_{pos}, s_{pos})$ and a negative edge $e_{neg} = n_{neg}$ \edge{NOT\_COLLECT $[a_{neg}]$} $(d_{neg}, s_{neg})$ in a \poligraph{} are conflicting if all the following parameters conflicts:

\begin{compactitem}[$\bullet$]
  \item \textbf{\textit{Data types}} $d_{pos}$ and $d_{neg}$ conflict \textit{iff} $d_{pos} = d_{neg}$ or $\exists d': subsume(d_{pos}, d') \land subsume(d_{neg}, d')$.
  \item \textbf{\textit{Entities}} $n_{pos}$ and $n_{neg}$ conflict \textit{iff} $n_{pos} = n_{neg}$ or $\exists n': subsume(n_{pos}, n') \land subsume(n_{neg}, n')$.
  \item \textbf{\textit{Purposes}} {\small $P_{pos}=Purposes(e_{pos})$ and $P_{neg}=Purposes(e_{neg})$} conflict \textit{iff} (1) $P_{pos} \cap P_{neg} \neq \varnothing$, or (2) $P_{neg} = \varnothing$.
  \item \textbf{\textit{Data subjects}} $s_{pos}$ and $s_{neg}$ conflict \textit{iff} $s_{pos} = s_{neg}$. 
  \item \textbf{\textit{Actions}} $a_{pos}$ and $a_{neg}$ conflict \textit{iff} $a_{pos} = a_{neg}$.    
\end{compactitem}
\label{def:contradicting}
\end{definition}

\subsubsection{Results on Contradiction Analysis}
\label{subsubsec:contradiciton-results}

\paragraph{Reducing False Alarms}
\poligraph{} with the above extensions encodes  additional parameters, which allows us to reclassify many of the statements erroneously classified as candidate contradictions in PolicyLint, due to missing context, as non-contradictions. To evaluate the benefit, (1) we map contradicting tuples reported by PolicyLint to \poligraph{} edges and (2) we check whether each pair of edges are conflicting as defined above.
\tool{} maps 2,555 PolicyLint contradictions to 1,566 pairs of edges in our dataset. Out of them, only 13.5\% (211) pairs are conflicting, taking contexts into account as per Definition~\ref{def:contradicting}. The remaining 86.5\% are not conflicting, due to one or more of the reasons shown in Table~\ref{tab:contradiction-reclassification}.
Please see~\Cref{appendix:contradiction-methodology} for details.

\paragraph{Validation}\label{subsec:validation-of-contraditions}
We manually verified all the 83 contradictions reported by PolicyLint in our test set. \tool{} reported 68 pairs as non-conflicting and we agree with all of them. For the other 15 pairs of conflicting edges, 7 pairs of edges are considered contradictions by human readers.

Despite the additional contexts captured by our extended framework compared to PolicyLint, \tool{} still has false alarms.
We manually verify all 211 pairs of conflicting edges identified by \tool{}, and find that only 25.1\% (53) pairs are real contradictions. The most common reason (54.0\%, 114 pairs) why conflicting edges may not be real contradictions is additional language nuances,  not yet represented in \poligraph{}.
For example, the sentence ``we do not collect personal data when you visit the site'' does not contradict with ``we collect personal data when you sign up'' due to the different conditions addressed. Other contexts we identify during manual validation include data sources and consent types.
Please see~\Cref{appendix:contradiction-validation} for details.

\begin{table}[t!]
\renewcommand{\arraystretch}{0.9}
\centering
\caption{Reclassification of PolicyLint contradictions}
\label{tab:contradiction-reclassification}
\small{
\begin{tabular}{@{}llrl@{}}
\toprule
\multicolumn{2}{@{}l}{}                                                             & \multicolumn{2}{l@{}}{\textbf{\# pairs of edges}} \\ \midrule
\multicolumn{2}{@{}l}{\textbf{Invalid*}}                                            & 183                 & (11.7\%)                 \\
\multicolumn{2}{@{}l}{\textbf{Non-conflicting parameters}}                          & 731                 & (46.7\%)                 \\
                         & \textit{Different purposes}                           & \textit{114}        & \textit{(7.3\%)}         \\
                         & \textit{Different data subjects}                      & \textit{121}        & \textit{(7.7\%)}         \\
                         & \textit{Different actions}                            & \textit{624}        & \textit{(39.8\%)}        \\
\multicolumn{2}{@{}l}{\textbf{Contradictions according to PolicyLint's ontologies}} & 441                 & (28.2\%)                 \\
\multicolumn{2}{@{}l}{\textbf{Conflicting edges}}                                   & 211                 & (13.5\%)                 \\ \midrule
\multicolumn{2}{@{}l}{\textbf{Total}}                                               & 1,566               &                          \\ \bottomrule
\multicolumn{4}{l}{\begin{tabular}[c]{@{}l@{}}* ``Invalid'' means that \tool{} maps a negative PolicyLint tuple to \\ a positive edge or the reverse. This is often because PolicyLint misinterprets\\
positive or negative sentences due to NLP limitations.\end{tabular}}
\end{tabular}
}

\end{table}

\paragraph{Conclusion} Our extensions of \poligraph{} to analyze contradictions, by taking account of more contexts, prevent many false alarms in prior work and narrow down possible contradictions to a smaller set.  However, one should be aware that even the extended framework does not cover all possible  contexts and nuances in human language. %

\subsection{Data Flow-to-Policy Consistency Analysis}
\label{sec:flow-to-policy-consistency-analysis}

In this section, we compare the statements made in a policy (extracted using \poligraph{}) to  the actual data collection practices (as observed in the network traffic generated). This application has been previously explored by PoliCheck for mobile apps~\cite{andow2020actions}  and its adaptations to other app ecosystems, \eg{} smart speakers~\cite{lentzsch2021hey,echoespaper} and VR devices~\cite{trimananda2022ovrseen}.

\subsubsection{Consistency Model}
\label{subsec:flow-to-policy-consistency-definition}

As in prior work~\cite{andow2020actions, bui2021consistency}, we represent data collection practices observed in the network traffic as data flows. A data flow is a tuple $f = (n, d)$ where $d$ is the data type that is sent to an entity $n$.
Given a \poligraph{}, we check whether the data flow is clearly disclosed in it as below.

\begin{definition}{\textbf{\textit{Clear Disclosure.}}}
\label{def:clear-disclosure}
\looseness=-1
Following \Cref{def:poligraph}, $G = \langle D, N; E_{S}, E_{C}; P \rangle$ is a \poligraph{}.
A data flow $f = (n, d)$ is \textit{clearly disclosed} in the policy represented by $G$ \textit{iff} it contains the entity ($n \in N$) and the data type ($d \in D$), and $collect(n, d)$ is true in $G$.
\end{definition}

Recall that PoliCheck also accepts broader terms of data types and entities as consistent but \textit{vague} disclosure, if the broader terms subsume the specific data types in the data flows according to the global ontologies. We define vague disclosures of a data flow in a similar way as follows.

\begin{definition}{\textbf{\textit{Vague Disclosure.}}}
\label{def:vague-disclosure}
Following \Cref{def:poligraph,def:global-ontology},
$G = \langle D, N;\allowbreak E_{S}, E_{C};\allowbreak P \rangle$ is a \poligraph{},
$O_d=\langle D_{d}, E_{d} \rangle$ is the global data ontology, and $O_n=\langle N_{n},E_{n} \rangle$ is the global entity ontology.
A data flow $f = (n, d)$ is \textit{vaguely disclosed} in $G$ according to global ontologies $O_d$ and $O_n$,
\textit{iff} $f$ is not clearly disclosed in $G$, and
there exist a data type $d' \in D \cap D_d$ and an entity $n' \in N \cap N_n$ that satisfy: $collect(n', d')$ is true in $G$; $subsume(d', d)$ is true in $O_d$; and $subsume(n', n)$ is true in $O_n$.
\end{definition}

Our definitions of clear and vague disclosures correspond to the same concepts in PoliCheck.
Both clear and vague disclosures are considered \textit{consistent disclosures}. Otherwise, the data flow has \textit{inconsistent disclosure} in the policy.

\subsubsection{Data Flow-to-Policy Consistency Results}
\label{appendix:flow-to-policy-result}

To facilitate a comparison to PoliCheck,
we use the same dataset from it~\cite{andow2020actions}. In addition to policy URLs for apps, this dataset also contains data flows extracted from the apps' network traffic.
Different from the policy dataset in \Cref{sec:evaluation}, we use the versions of policies from 2019 for a fair comparison with PoliCheck results, because the data flows were collected around February 2019.
We crawl the historical versions of policies from Internet Archive~\cite{internet-archive}.
In total, we have 8,757 apps with both data flows and policies available.

\begin{figure*}[!t]
\minipage{0.47\textwidth}
    \includegraphics[width=\columnwidth,page=1]{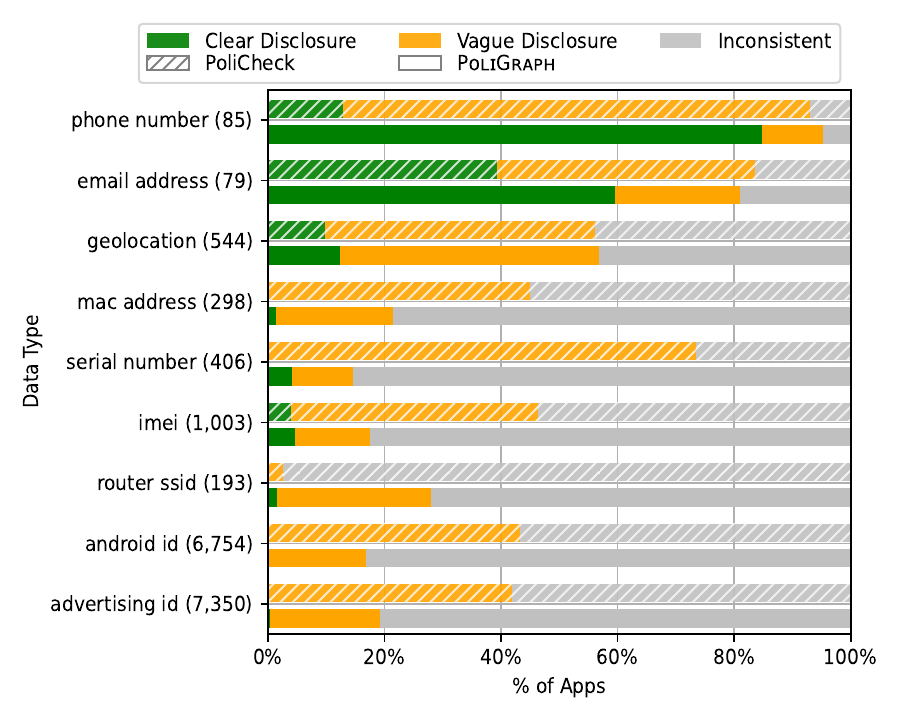}
    \caption{\textbf{Flow-to-policy consistency comparison of \poligraph{} \vs{} PoliCheck.}
    The numbers in parentheses represent the number of apps that collect the data type. For example, 85 apps have data flows that collect ``phone number''.
    }
    \label{fig:network-poligraph}
\endminipage\hfill
\minipage{0.52\textwidth}
    \vspace{1em}
    \includegraphics[width=\columnwidth,page=2]{figures/result-network.pdf}
    \caption{\textbf{Purposes of data flows inferred from \poligraph{}s.}Some apps disclose the collection of the data types for \textit{core} or \textit{non-core} functionality, or both. We ignore apps whose policies do not disclose purposes, so the total number of apps (listed in parentheses) is less than the number of apps in Figure~\ref{fig:network-poligraph}.}
    \label{fig:network-purpose}
\endminipage
\end{figure*}

\Cref{fig:network-poligraph} compares flow-to-policy consistency results per data type in  \poligraph{} \vs{} PoliCheck.
An app may send one data type to multiple entities, resulting in multiple data flows per data type. In this case, we report the worst disclosure type for the app, \eg{} if, at least, one of the data flows of the data type is inconsistent, the disclosure type is reported as inconsistent. 
We present the results for nine data types---three out of 12 data types are analyzed in~\cite{andow2020actions}, but excluded here as only less than 10 apps exhibit the three in their data flows.

\paragraph{Clear  disclosures} %
We find that \tool{} outperforms  PoliCheck in terms of capturing more clear disclosures. PoliCheck underestimates the number of clear disclosures for all data types due to its limited recall (see \Cref{subsec:comparison-with-policylint}).
Clear disclosure of contact information is especially underestimated.
72 of 85 apps that collect ``phone number'', and 47 of 79 apps that collect ``email address'' clearly disclose the collection.

\paragraph{Vague disclosures} %
Here, \tool{} extracts fewer vague disclosures than PoliCheck. Further investigation reveals that this is because our global data ontology has a different design compared to PoliCheck's. In PoliCheck's data ontology, ``personal information'', a commonly seen term in policies, subsumes ``device identifiers''. While this would effectively increase PoliCheck's coverage, namely that the collection of all data types related to ``device identifiers'' found in the data flows would be categorized as vague disclosures, it is unclear whether ``device identifiers'' can be strictly categorized as ``personal information''.
Many policies do not consider ``device identifiers'' as ``personal information''.
\revision{On the contrary, for ``router ssid'', \tool{} actually found more vague disclosures. It is because PoliCheck's data ontology %
 does not count it as ``device identifier'', while we do so. The results indicate that the number of vague disclosures are largely affected by ontologies.}

\revision{
\subsubsection{Purposes of Data Flows}
\poligraph{} also captures the purposes of data collection, use and sharing. This enables us to enrich the flow-to-policy consistency analysis by identifying the purposes associated with data flows according to policies.

Inferring the purposes from the network traffic would be limited to few purposes~\cite{jin2018mobipurpose}. For example, while \textit{advertising} as a purpose can be inferred based on a network packet's destination entity, it would not be straightforward to infer \textit{legal} as a purpose,  and such inferences often do not  align well with the definitions given in the policies~\cite{bui2021consistency}.
Thus, instead of directly checking the consistency between the purposes disclosed in the policies and the data flows in the network traffic as in prior work~\cite{bui2021consistency}, we adopt the methodology used in prior work~\cite{trimananda2022ovrseen} that infers the purposes of each data flow based on collection statements in policies that disclose purposes. This is captured in the corresponding \poligraph{} as a \COLLECT{} edge assigned with a list of attributes that contain purposes (see Section~\ref{sec:knowledge-graph}).

Following Definitions~\ref{def:clear-disclosure} and~\ref{def:vague-disclosure} in Section~\ref{subsec:flow-to-policy-consistency-definition}, we extract the set of purposes $\mathit{purposes}(n, d)$ (see Definition~\ref{def:set-of-purposes}) for each clear disclosure, or $\mathit{purposes}(n', d')$ for each vague disclosure, as the inferred purposes of the corresponding data flow $(n, d)$.
We are able to extract purposes for clearly and vaguely disclosed data flows of 1,023 apps based on \poligraph{}s of their policies.
We label the data flows with the five categories of purposes, and further group them into core and non-core purposes as discussed in Section~\ref{sec:building-knowledge-graph}. Services, security, and legal are grouped as core purposes; whereas analytics and advertising are grouped as non-core purposes.

Figure~\ref{fig:network-purpose} shows the purposes associated with these data flows.
In total, 707 apps declare that their data flows are for core purposes, whereas 692 apps declare that their data flows are for non-core purposes.
While the collection of ``advertising id'' (and other similar identifiers) for non-core purposes (\ie{} advertising and analytics) can be acceptable, Figure~\ref{fig:network-purpose} also shows that many apps collect sensitive data types, \eg{} ``phone number'', ``email address'', and ``geolocation'' for non-core purposes, which is alarming and aligns with Finding~\ref{finding:non-core-usage} in Section~\ref{sec:summarization}.
Note that many policies declare data flows for both core and non-core purposes. This is usually due to vague disclosures and multiple purposes associated with generic terms, which reduces transparency. A similar issue about data sharing disclosed in generic terms is discussed in Finding~\ref{generic-term} in Section~\ref{sec:summarization}.
}

\section{\llmtool{}: Extracting \poligraph{} using Large Language Models}
\label{sec:llmtool}

After we developed \tool{}, large language models (LLMs)%
\footnote{We use the term ``LLMs'' to refer to GPT-like {\em text generation models} that take text prompts as input and generate text as output. We do not refer to other task-specific language models (\eg{} BERT-based classification models~\cite{liu2019roberta}) or large models that mainly work with other modalities.} %
represented by ChatGPT~\cite{chatgpt-web,openai-gpt4o-mini,openai-gpt4o-system-card} emerged and greatly enhanced the capabilities of natural language processing (NLP).
LLMs are able to understand complex human language (\eg{} user queries and any human written articles), and generate high-quality natural language responses.
More importantly, LLMs provide a generic natural language interface that can be used to do many tasks in a unified and zero- (or few-) shot way, thus have the potential to largely replace linguistic analysis, task-specific training and heuristics.
In this section, we explore the alternative implementation of \tool{} using LLMs.

\subsection{LLM for Privacy Policy Analysis: Challenges and Solutions}
\label{sec:llmtool-challenges-and-solutions}

In a naive use of LLMs, one can input a policy into an LLM and ask it to answer questions or to extract relevant information. There are still technical challenges to be addressed.

First, the natural language interface of LLM is intended for human interaction and does not naturally lend itself to automation tasks.
In our application scenario, we intend to extract information about data collection from policies in a {\em machine-readable} format that is used for further automated analysis (\eg{} auditing data collection, see \Cref{sec:flow-to-policy-consistency-analysis}) with non-machine-learning components.
In this regard, we believe that an analysis framework, namely \poligraph{}, is still relevant and necessary to guide LLM analysis and define the interface for automated analysis.

Second, LLMs have known limitations, such as limited context length, hallucination (\ie{} generating irrelevant responses), and coverage errors (\ie{} omitting relevant information)~\cite{chang-etal-2024-detecting}.
At least at the moment this paper is written, it is not practical to direct LLMs to extract complete and accurate information (\eg{} a full \poligraph{}, or all PoliCheck tuples) from lengthy policy text in an end-to-end manner.
In order to divide the whole task into manageable subtasks, and implement heuristics to verify LLM outputs, we use the modular design of \tool{} as a good start.

In this section, we present \llmtool, an LLM-based policy analyzer and a re-implementation of \tool{} that extracts \poligraph{}s through LLM prompting instead of linguistic analysis.
\llmtool{} has two main LLM-based components:
\begin{inenum}
    \item an annotator module that extracts parameters, or phrases, of interest (\eg{} data types, recipient entities, purposes \etc) from the policy text, and
    \item several phrase normalization modules that maps extracted parameters into standard terms and build local ontologies, allowing automated analysis.
\end{inenum}
Our evaluation shows that \llmtool{} extracts \poligraph{}s with high precision and significantly better recall than the previous linguistic-analysis-based method, namely~\tool{}.

\subsection{Design Choices}
\label{sec:llmpg-design-decisions}

Before we dive into \llmtool{}'s implementation details, we highlight the following high-level design choices.

\paragraph{Model Choice}
We anticipate that LLMs' capabilities will continue to increase rapidly.
Therefore, we would like our methodology to be model-agnostic and reusable in the future with more advanced LLMs.
We use OpenAI's GPT 4o-mini model for evaluating \llmtool{} in~\Cref{sec:llmpg-evaluation} due to its great cost efficiency.
However, any state-of-the-art LLM that supports at least thousands of tokens in its context window, such as Meta's Llama 3~\cite{meta-llama3} and OpenAI GPT-4o series~\cite{openai-gpt4o-system-card,openai-gpt4o-mini}.
We simply refer to the model as {\em the LLM} in this section.

\paragraph{Prompt Engineering}
There are two main techniques to adapt LLMs for domain-specific task: {\em fine-tuning}~\cite{openai-api-fine-tuning} and {\em prompt engineering}~\cite{openai-api-prompt-engineering}.
Fine-tuning involves training the LLM with extra domain- or task-specific corpus to align its outputs better with the ground truth.
Prompt engineering, assuming the LLM have had the knowledge for the task, focuses on crafting the instructions given to the LLM.
We choose prompt engineering to benefit from LLMs' existing capabilities, \eg{} language understanding, world knowledge and commonsense reasoning.
Our intuition is that understanding policies requires no more than these capabilities.
Recent LLMs support sufficiently long input (\eg{} from 8K to more than 100K input tokens~\cite{meta-llama3,openai-gpt4o-system-card,openai-gpt4o-mini}), which allows us to describe our tasks in great details.
In contrast, fine-tuning requires significant human labor to create the training corpus and computational resources for training, and the fine-tuned models are often soon outpaced by newer general LLMs.
For example, \citet{chanenson2023automating} fine-tuned GPT-3.5 for annotating policies, which has been deprecated by OpenAI one year after the work~\cite{openai-api-deprecations}.

\paragraph{Constraining LLM Output}
It is well known, as discussed earlier, that LLMs can hallucinate.
Let's consider, for example, prompting LLMs to extract
personal data types from the policy.
Hallucination means that the LLM may generate inaccurate or bogus output \eg{} a data type that is never mentioned.
More subtly, the LLM occasionally fails to follow the prompt strictly, \eg{} it can output in a different format that causes trouble for automation.
Hallucination hinders the {\em precision} of the task as the extracted information can be irrelevant and undesirable in our application.

Our approach to mitigating hallucination is to apply non-machine-learning heuristics to validate and constrain LLM output.
For example, in the \llmtool{} annotator, we programmatically verify that the extracted phrases are indeed from the given policy text, and discard those that are not.
Constrained or guided LLM generation is a well-studied technique~\cite{bk-prompting,zheng2023efficiently}.
Since we cannot implement token-level constraints through commercial LLM APIs, we instead heuristically verify and correct the output after LLM generation in \llmtool{}.

\paragraph{LLM Reflection}
On the other hand, the coverage errors of LLMs,
\ie{} omitting legitimate results (see~\Cref{sec:llmtool-challenges-and-solutions}), hinder the {\em recall} of the task.
Coverage errors are especially an issue when there is a lot of information to be extracted in lengthy and complicated text, such as in a policy.
A low coverage would offset the benefit of using LLMs to understand more text than non-LLM methods (\eg{} linguistic analysis). 

To mitigate coverage errors, we iteratively instruct the LLM to analyze its own output, assess the completeness, and update the output if necessary.
This strategy is known as reflection~\cite{llm-reflection}.
Reflection can also help mitigate hallucination by allowing the LLM to make corrections in the updated output.

\begin{figure}[t!]
    \centering
    \includegraphics[width=\linewidth]{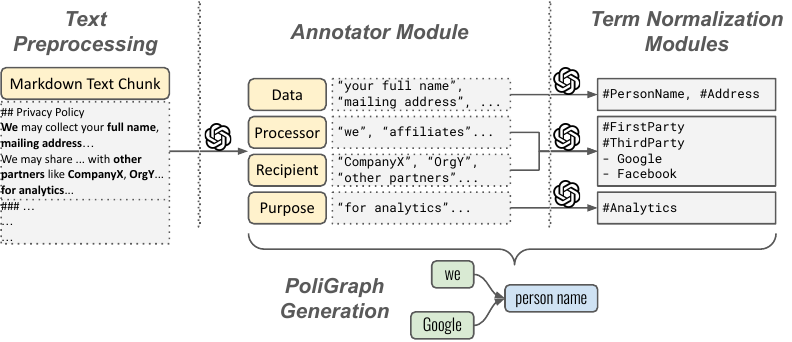}
    \caption{\textbf{Overview of \llmtool{} implementation.} First, \llmtool{} preprocesses the HTML-format policy into Markdown format and split the text into overlapping chunks.
    Second, the LLM-based annotator module extracts data collection statements and represent them using a predefined list of parameters.
    Then, the LLM-based term normalization modules maps data types, entities and purposes into standard terms.
    Finally, all the LLM extracted information is combined to generate a \poligraph{}.}
    \label{fig:llmtool-overview}
\end{figure}

\subsection{\llmtool{} Implementation}
\label{sec:llm-implementation}

We present \llmtool, the LLM-based system that we build to generate the \poligraph{} from the text of a policy.
\Cref{fig:llmtool-overview} gives an overview of the implementation.

\subsubsection{Text Preprocessing}

\llmtool{} simply processes a policy, originally in HTML format, into Markdown format, which is able to preserve common text structures, including headings, lists, and tables.
Unlike linguistic analysis, LLMs do not only recognize complete sentences, but also understand fragmented text in lists, tables and other structures.

As many LLMs have limited context length or do not work well on long text~\cite{li2024long}, we also segment the Markdown text into overlapping {\em text chunks}.
A text chunk is the smallest unit that LLM-based modules process.
We set the chunk size to 500 tokens.%
\footnote{While recent LLMs support much longer input, they do not necessary perform well on long text~\cite{li2024long}.
Moreover, LLMs often have a limit on the number of output tokens that is around 4K to 16K~\cite{meta-llama3,openai-gpt4o-mini}.
In consideration of these limits, we do not use larger chunk size.
Moreover, larger chunks would make it more difficult to programmatically associate the LLM's output with the original text.
}
To mitigate possible truncated context due to segmentation, we let consecutive text chunks overlap, and the segmentation never happens in the middle of a paragraph, list item, or table.
We also include headings at the beginning of each segment to provide additional context.

\subsubsection{Annotator Module}
\label{sec:llm-annotator-module}

In \llmtool{}, the annotator is an LLM-based module that identifies statements about data collection, use and sharing (\ie{} data collection statements) in the policy, and extract various parameters from such statements, just like annotators in \tool{}.
However, in contrast to the rule-based linguistic analysis in \tool{} that parses the dependency tree of a sentence, the annotator in \llmtool{} reads an entire chunk of text and determines the {\em semantic roles} of relevant words and phrases.
Semantic roles break down the text into parameters that describe, for example, ``who'' (\eg{} the first party), ``did what'' (\eg{} sharing), ``to whom'' (\eg{} third parties), ``for what purpose'' (\eg{} for advertising) and ``in what context'' (\eg{} when users opt in).
This is closer to how human understands the text than traditional NLP linguistic analysis.

Specifically, the annotator takes a chunk of policy text, and prompts the LLM to extract the following parameters if the text contains any data collection statements:
\begin{inenum}
    \item {\em data types} being processed, such as ``personal data'', ``email address'';
    \item {\em actions} applied to the data types, such as ``collect'', ``use'', ``share'';
    \item {\em processors}, \ie{} entities that do the action on the data;
    \item {\em recipients}, \ie{} entities that receive the data, when the action involves data transfer;
    \item {\em purposes} of data processing, such as ``for authentication'', ``in order to provide ads''; and
    \item {\em contexts}, \ie{} other conditions associated with data processing, such as ``if you register an account'', ``when you use the service''.
    Additionally, to handle negative statements, we add another parameter:
    \item {\em prohibition}, a boolean value that, if set to true, indicates the stated action is prohibited or denied.
\end{inenum}
Note that these parameters are chosen to describe semantic roles and do not map directly to \poligraph{} components yet.

In the prompt for the annotator module, we describe the task, as well as including several examples for {\em few-shot learning} in the prompt.
We also instruct the LLM to output extracted parameters as JSON objects and quotes the original phrases in the policy text.
This allows the annotator to programmatically verify the extracted parameters and discard those that do not appear in the text (\ie{} hallucinations; see \Cref{sec:llmpg-design-decisions}).
The full LLM prompts for the annotator module is provided in \Cref{appendix:llm-prompt-annotator}.
In practice, the LLM still makes minor mistakes occasionally, \eg{} stripping the plural suffix.
To avoid discarding too many results, the annotator performs fuzzy string matching~\cite{github-thefuzz} and attempts to correct minor differences in the LLM output.

To improve the coverage, the annotator performs 3 rounds of reflection in \llmtool{}.
In each round, the LLM is instructed to review the text and the parameters extracted so far, and asked if there are more to be added.

\subsubsection{Term Normalization Modules}
\label{sec:llm-term-norm-implementation}

To facilitate automated analysis, \llmtool{} maps / normalizes extracted parameters of data collection statements to standard terms in the \poligraph{} ontologies.
This is the same function as in \tool{} (see \Cref{sec:building-knowledge-graph}) but implemented by prompting the LLM.
We implement LLM-based term normalization modules for data types, partly for recipient entities (\ie{} processors and recipients), and for purposes.

\paragraph{Data Types}
The data type normalization module provides the policy text (in chunks) and \poligraph{}'s data type ontology (including the hierarchy and the definitions of the terms, see \Cref{sec:ontology}) to the LLM, asking it to map phrases of data types to ontology terms if possible.
In the text, data type phrases (as identified by the annotator) is quoted using a specific format (\eg{} ``we collect \{contact details\}'') to indicate the phrases to process.
The normalization results are outputted as key-value pairs in a JSON object (\eg{} \path{"contact details": {"concept": "#ContactInfo"}}).
If a phrase does not match any standard term, a special value \path{#Other} is expected.

The module also instructs the LLM to resolve references between phrases, including co-reference and subsumption.
For example, in the sentence ``we collect technical information such as your IP address, session ID, and ...'',
``technical information'' does not match any standard term in the ontology, but it refers to (or subsumes) other terms.
The module instructs the LLM to return the referenced phrases (\ie{}
\path{"contact details": {"referents": ["your IP address", "session ID", ...]}}).
The information is used to create subsumption edges in \poligraph{}.
This replaces the subsumption annotator and coreference annotator in \tool{} (see \Cref{sec:annotators}).
The full LLM prompt for the data type normalization module is provided in \Cref{appendix:llm-prompt-data-type-normalization}.

The normalization results are verified programmatically.
Any terms not found in the ontology, and any phrases not found in the text, are discarded to prevent hallucination.
The module implements five rounds of reflection for data type normalization as follows.
First, it instructs the LLM to include new phrases of data types in the results if they are referred by any quoted data types.
This helps to cover more data types  omitted by the annotator.
Second, the LLM may discard any quoted phrases if it determines they do not refer to data types, allowing to fix mistakes by the annotator.

\paragraph{Entities}
\looseness=-1
For processor and recipient entities, it is impossible provide the entire company list in the prompt.
In \llmtool{}'s entity normalization module, we only prompt the LLM to classify entities to either first party (\eg{} ``we'' or the name of the developer or vendor), third party (\ie{} other data processors or recipients), or other (\eg{} the user).
We reuse the regular-expression-based phrase normalization in \tool{} (see \Cref{sec:building-knowledge-graph}) to further map third-party entities to ontology nodes and specific company names.
The full LLM prompt for the data type normalization module is provided in \Cref{appendix:llm-prompt-entity-normalization}.

\paragraph{Purposes}
The purpose normalization module prompts the LLM to classify each purpose phrase extracted by the annotator into one or more of the six categories:
\textit{services},
\textit{security},
\textit{legal},
\textit{advertising},
\textit{analytics},
and \textit{other} (\ie{} any other purposes that do not fit into previous categories).
The categories are consistent with \tool{} for compatibility.
However, we do note that it is possible to do finer-grained purpose classification with LLMs.

\subsubsection{\poligraph{} Generation}
\llmtool{} combines parameters of data collection statements from all different chunks to create a \poligraph{}.
For each collection statement, the data type is considered collected by the recipient entity, or the processor, depending on the action (\eg{} collect, or share), and they are connected by \COLLECT{} or \NOTCOLLECT{} edges depending on the prohibition parameter.
\COLLECT{} and \NOTCOLLECT{} edges are annotated with the purposes of the data collection if available.
\SUBSUME{} edges are created based on the referents and subsumption relations identified by the normalization modules\footnote{In our current implementation, \llmtool{} only fully resolves referents and subsumption relations for data types. Entities are classified into first and third parties. So for entities, \SUBSUME{} edges only exists between each entity and either ``we'' or ``third party'' node.}.

Note that \llmtool{} does not yet incorporate all the information extracted by the annotator (\eg{} context, action, and distinguishing between processors and recipients).
We leave these for future extension of \poligraph{} or for use with other more sophisticated knowledge graph structures (\eg{} Data Privacy Vocabulary~\cite{dpv2_0}).

\subsection{Evaluation}
\label{sec:llmpg-evaluation}

In this section, we evaluate \llmtool{}'s performance in analyzing policies.
For the evaluation, we use OpenAI's GPT 4o-mini model~\cite{openai-gpt4o-mini} as the LLM implementation.
Due to the high cost of running LLM (see~\Cref{sec:llm-cost} for more analysis),  we only evaluate \llmtool{} with our test set -- 200 policies from the PoliCheck dataset (see \Cref{sec:evaluation}).

\subsubsection{Comparison to Other Implementations}
\label{subsec:llmpg-comparison}

In order to compare \llmtool{} with \tool{} and prior work, we repeat the evaluation in \Cref{subsec:comparison-with-policylint}.
That is, we validate the collection relations (\ie{} $collect(n, d)$) in \poligraph{}s against manually labeled ground truth (for a subset of data types, and only considering whether the entity is first or third party).

\paragraph{Baselines}
We compare \llmtool{} with \tool{} and a naive LLM privacy policy analyzer implementation.
The naive analyzer extracts tuples \textit{(first/third party, data type)} by simply prompting the LLM (GPT-4o mini) to read the policy text and output whether each data type is collected by the first or third party, in an end-to-end manner.
It does not include any programmatical verification or reflection steps.
Note this implementation is only meant for evaluation purposes and is by no means comparable to \llmtool{} in terms of functionality -- it only queries for specific data types and entities that we consider for this evaluation.
Please see \Cref{appendix:llm-prompt-naive-analyzer} for the prompt used.

\Cref{tab:llmtool-evaluation} presents the evaluation results.
First, compared with \tool{}, \llmtool{} further improves the recall of collection relations, from 66\% to 83\%, while maintaining a good precision of 93\%.
Note that the extra relations discovered by \llmtool{}, most of which are third-party ones, result in a higher number of tuples and a lower recall for \tool{} compared with \Cref{subsec:comparison-with-policylint}.
Then, the naive LLM analyzer, albeit with a higher recall, has a much lower precision than \llmtool{}.
This is because it tends to hallucinate a lot by overly interpreting the vague language -- despite that we have instructed the LLM to only include explicitly mentioned data types. %
This highlights the importance of output constraining and reflection steps in \llmtool{}.

\begin{table}[t!]
\centering
\caption{Manual validation of \llmtool{} results.}
\label{tab:llmtool-evaluation}
\centering
\begin{tabular}{@{}l@{}cll@{}}
\toprule
& \textbf{\#tuples} & \textbf{prec. (1st/3rd party)} & \textbf{recall (1st/3rd party)} \\ \midrule
Ground Truth                        &   950 & -                        & -                        \\
\tool{}                           &   640 & 97.2\% (99.8\% / 92.8\%) & 65.5\% (62.8\% / 71.1\%) \\
Naive LLM Analyzer                  & 1,414 & 65.3\% (77.5\% / 49.8\%) & 92.7\% (95.5\% / 87.6\%) \\
\textbf{\llmtool{}}                 &   893 & 92.8\% (95.7\% / 88.1\%) & 83.2\% (83.0\% / 83.6\%) \\
\bottomrule
\end{tabular}
\end{table}

\paragraph{Recall}
The improvement in recall can be largely attributed to the exceptional language understanding capabilities of the LLM.
Unlike previous rule-based approaches, the LLM is able to grasp the meaning of more complex sentences.
For example, consider the following sentence from a policy:
``We may also provide a more typical registration flow where you may be required to provide ... your age and birthday.''\cite{hc_games_privacy_policy}
The sentence is not recognized by \tool{} because its rules do not cover the adverb clause.
Moreover, \llmtool{} also understands information in Markdown tables that are not complete sentences. 
In some instances, \llmtool{} is able to identify data collection statements that were subtly implied and even previously overlooked by human readers.

\paragraph{Precision}
\llmtool{} achieves a notably lower precision on our test set.
However, this is largely due to disagreements between human and LLM's interpretation of the vague policy language.
For example, the LLM normalizes some phrases differently than humans.
In one policy~\cite{fingerfun_privacy_policy}, \llmtool{} maps ``mobile device unique device id'' to ``android id''.
This is not completely wrong but we reject it because \tool's regular-expression-based term normalization did not include such keywords.
In a more subtle case, a policy states
``As part of the registration process for our e-newsletters, we collect personal information... We use a third-party provider, MailChimp, to deliver our newsletter''\cite{aimer_media_privacy_policy}, and the LLM identifies that ``MailChimp'' may collect ``personal information''.
We reject this because the two sentences are not directly related by any common data type or entity phrases.
On the one hand, the misalignment shows the limitation of hard-label ground truth (\ie{} every collection relation has to be labeled either true or false) in the face of vague language.
On the other hand, unlike rule-based approaches or finetuning, it is hard to precisely align the LLM's interpretation of the text with human understanding.

\subsubsection{Cost Analysis}
\label{sec:llm-cost}

A main obstacle to use LLMs at scale is the cost.
Unlike \tool{} which uses lightweight RoBERTa models (see~\Cref{sec:implementation}) and can run on a personal computer, the LLM that support \llmtool{} cannot be easily run locally.
In fact, many state-of-the-art LLMs, include the GPT 4o-mini model we used, are only accessible through commercial APIs that incurs monetary cost.
On our test set, \llmtool{} uses about 110K input tokens per policy.
To control the cost, we use GPT-4o mini~\cite{openai-gpt4o-mini}, one of the most cost-efficient LLMs, which costs about 0.017 USD per policy.
The cost could be prohibitive if more advanced models were used.
For example, using GPT-4o~\cite{openai-gpt4o-system-card}, one of the most advanced LLMs, the cost would be about 0.55 USD per policy.
However, with the expected decreasing cost of LLMs, we expect the cost not to be prohibitive in the long run.

Due to the high cost of LLMs, we only applied \llmtool{} to analyze our test set and did not repeat all the analysis and applications in~\Cref{sec:evaluation,sec:applications}. Note that the ablation study in~\Cref{subsec:ablation-study} also does not apply because the LLM-based modules work in a pipeline and cannot be disabled separately. In \Cref{appendix:llm-component-eval}, we evaluate the effectiveness of reflection in the annotator module and the precision of the normalization modules.

\section{Conclusion}
\label{sec:poligraph-discussion}

\subsection{Summary}

We present \poligraph{}, a framework that represents a privacy policy as a knowledge graph that
\begin{inenum}
    \item connects statements about data collection, use and sharing across different parts of the policy text, and 
    \item clearly defines and distinguishes between local and global ontologies.
\end{inenum}
We design and implement \tool{}---a tool that leverages NLP linguistic analysis to generate \poligraph{}s from policy text.
Because \poligraph{} allows inferences of collection relations across paragraphs and sections, it significantly improves {\em recall} over prior work, while maintaining a high {\em precision} (see \Cref{subsec:comparison-with-policylint}).
Our manual validation shows that \tool{} improves the recall of collection statements from 30\% to 70\% in comparison to PolicyLint~\cite{andow2019policylint}. Meanwhile, the precision is improved from 92\% to 97\%.
\poligraph{}, and the global ontologies used with it, also enable new privacy policy analyses that were not previously possible: summarizing patterns in a corpus of privacy policies (see \Cref{sec:summarization}) and assessing the correctness of definitions of terms (see \Cref{sec:terms-and-definitions}). 
Our modular design also allows for extensions to the framework so as to analyze contradicting statements in more fine-grained contexts than prior work (see \Cref{sec:negative-statements}).
Last but not least, we present \llmtool{}, an LLM-based reimplementation of \tool{} that leverages the LLM to analyze \privacypolicy{} text, extract relevant information, and generate \poligraph{}s.
Our evaluation shows that \llmtool{} improves the recall of data collection statements from 60\% to 70\% compared to \tool{}.

\revision{
\subsection{Future Directions}

We anticipate the following trends and  directions in the  space of privacy policy analysis.

\paragraph{End-to-end Use of LLMs for Privacy Policy Analysis.}
LLMs' limitations, such as limited context length, hallucination, and coverage errors, prevent us from using an LLM in an end-to-end manner in \llmtool{} (see~\Cref{sec:llmtool}).
However, these limitations are not specific to our tasks, and we have seen progress in addressing them.
Recent work showed that newer and larger models are considerably less prone to hallucinate~\cite{li-etal-2023-halueval}.
In the future, we anticipate that more powerful LLMs will eliminate the need to decompose high-level tasks.
One can perform the desired analysis (\eg{} checking the consistency between network traffic and the policy) simply by supplementing the data and describing the task in the prompts.
Another obstacle in using LLMs at this moment is their high cost (monetary or computational).
In this regard, the LLM community is also creating smaller and more efficient models like Meta's Llama~\cite{meta-llama3} and OpenAI's GPT-4o mini~\cite{openai-gpt4o-mini}.
While smaller models do not yet match the capabilities of larger ones (\eg{} GPT-4o~\cite{openai-gpt4o-system-card}), we expect continued progress that will make LLMs increasingly capable and cost-effective.

\paragraph{Alternative Representations of Privacy Policies}
NLP analysis does not solve the problem of inherent vagueness and ambiguous language of policies.
The community proposes alternative representations of policies to mitigate the problem.
For example, app platforms such as as Apple's App Store and Google Play have introduced privacy labels or icons that provide concise summaries of policies~\cite{apple-privacy-details,google-data-safety-section}.
Similarly to ``nutrition labels'' for food, these labels enhance readability and reduce ambiguity by using predefined formats.
We anticipate their growing adoption in more platforms, especially on platforms where displaying long text is impractical, such as smart TVs and VR headsets.
However, different companies have designed these labels in different ways, hindering interoperability.
We hope that governments and academics lead the efforts to further standardize the format and the vocabulary of privacy labels.

\paragraph{Machine-Readable Privacy Policies}
Another way to eliminate the vagueness and ambiguity of human language is through machine-readable format.
In the past, the W3C P3P standard~\cite{w3c-p3p} proposed an XML schema to describe policies.
The Data Privacy Vocabulary (DPV) defines a knowledge graph scheme to express various aspects of policies,
including data processing, data rights, security measures and so on, in a machine-readable way.
DPV also defines a vocabulary of standard terms that facilitate interoperable representation of policies.
DPV can be seen as a superset of \poligraph{}.
In our work, we have originally used \poligraph{} to connect information from different sentences in a policy and to facilitate automation.
While a capable LLM in the future may be able to skip such representations to perform end tasks, machine-readable representations are useful in themselves.
They help eliminate ambiguity in human language and provide interoperability between LLM-based and non-machine-learning components in a system.
Going forward, an increasing use of machine-readable policies will facilitate the checking and enforcement of policies and laws automatically and at scale.
}

\begin{acks}
This work was supported by the National Science Foundation (Awards 1956393 and 1900654) and a gift from the UC Noyce Initiative. \revision{We would also like to thank Hamza Harkous for  discussions on LLMs and privacy policy analysis.}
\end{acks}

\bibliographystyle{ACM-Reference-Format}
\bibliography{reference,online}

{\noindent\Huge\bf Appendices}
\appendix
\section{\tool{} Implementation in Detail}
\label{appendix:implementation-details}

In Section~\ref{sec:implementation}, we present \tool{}, the NLP-based system that we implement to generate \poligraph{}. We provide additional details on its implementation in this appendix.

\subsection{HTML Preprocessing}
\label{appendix:html-preprocessing}

In Section~\ref{sec:nlp-on-structured-document}, we explain that \tool{} converts each policy into a simplified document tree structure for subsequent NLP tasks.
We provide more details in this appendix.

\paragraph{Document Tree} Given a policy in the HTML format, \tool{} starts by converting it into a \textit{document tree}, a simplified HTML DOM tree. Each node in the document tree corresponds to a fragment of text, referred to as a \textit{segment}.
\tool{} uses three kinds of segments: (1) A \textit{heading segment} corresponding to an HTML heading, whose parent, if exists, is the higher level heading segment; (2) A \textit{list-item segment} corresponding to an item in an HTML list, whose parent is the text or heading segment immediately before the list;  (3) A \textit{text segment} corresponding to a general HTML text container that is neither a heading nor a list item, whose parent is the heading closest to it.

A hypothetical document tree, modified from the policy example in Figure~\ref{fig:policy-example}(a), looks like this:

\begin{exmp}
\label{exmp:document-tree}
\textbf{\textit{Document Tree}}\\
\mbox{\texttt{~\colorbox{gray!20}{HEADING}}} Data Collection\\
\mbox{\texttt{~~~\colorbox{gray!20}{TEXT}}} We collect the following personal information:\\
\mbox{\texttt{~~~~~\colorbox{gray!20}{LISTITEM}}} - Device information, such as IP address...\\
\mbox{\texttt{~~~~~\colorbox{gray!20}{LISTITEM}}} - Location\\
\mbox{\texttt{~~~\colorbox{gray!20}{TEXT}}} We disclose the personal information as follows...
\end{exmp}

Technically, \tool{} first uses the Readability.js library~\cite{readability-js} to extract the main content of a webpage (\ie{} without sidebar, footer, and other unrelated widgets). Then it parses the HTML markups to generate the document tree. In some cases, webpages use plain text lists (\eg{} prefixing each paragraph with numbers) instead of HTML lists. \tool{} has heuristics to identify such lists and convert them into list segments.

\paragraph{Text Concatenation} NLP pipeline expects text input. And in order for it to produce correct results, we need to input complete sentences. Some segments, like the two list items in Example \ref{exmp:document-tree}, are not self-contained to form full sentences. Each of them has to be concatenated after its parent text segment to form a complete sentence, \eg{} ``We collect the following personal information: Name''. To do this, \tool{} uses the document tree to find the context of each segment.

In reality, nested headings and lists are common, and many policies are not written in a way that sentence boundaries can be easily identified, \ie{} it is hard to determine at which level a complete sentence starts without another NLP model. To work around the issue, \tool{} iteratively concatenates each segment with its parents up to different levels to get multiple versions of input text. For each \texttt{LISTITEM} in Example \ref{exmp:document-tree}, \tool{} generates three versions of the text for NLP:
(1) \texttt{LISTITEM};
(2) \texttt{TEXT+LISTITEM}; and
(3) \texttt{HEADING+TEXT+LISTITEM}.
NLP pipeline gets valid input as long as any of these versions aligns with sentence boundaries.

\paragraph{NLP Pipeline} \tool{} uses spaCy library~\cite{spacy3} and its \textit{en\_core\_web\_trf} NLP pipeline, which is a set of NLP models built upon the RoBERTa architecture~\cite{liu2019roberta}, to annotate common linguistic features. These linguistic features, including word lemmas, part-of-speech, sentence segmentation, and syntactic dependencies, are syntactic and, thus, they require no domain adaption. The RoBERTa-based pipeline achieves state-of-the-art performance on these tasks~\cite{spacy-trf-model}. We only train our own NER model for identifying data types and entities (see Appendix~\ref{appendix:ner-methodology}).
\tool{} relies on these linguistic features to build \poligraph{}s. Although spaCy's NLP pipeline processes plain text and does not keep the document tree structure, \tool{} internally records which segment each word comes from. This way, \tool{} can revert to the document tree whenever needed. For example, the list annotator (see Section~\ref{sec:annotators}) takes advantage of the document tree to discover additional relations.

\subsection{NER Model Training}
\label{appendix:ner-methodology}

\begin{table}[t!]
\centering
\caption{Root words used for rule-based NER}
\label{tab:ner-root-word}
\small{
\begin{tabular}{@{}ll@{}}
\toprule
\textbf{NER label} & \textbf{Root words of noun phrases} \\ \midrule
\textit{DATA}    & \begin{tabular}[c]{@{}p{6.5cm}@{}}
information, data, datum, address, number, location, geolocation, identifier, id, preference, setting, cookie
\end{tabular} \\ \midrule
\textit{ENTITY}      & \begin{tabular}[c]{@{}p{6.5cm}@{}}
agency, advertiser, affiliate, analytic, analytics, app, application, broker, business, carrier, company, corporation, distributor, group, institution, network, operator, organization, partner, party, platform, processor, product, provider, publisher, service, site, software, subsidiary, vendor, website
\end{tabular} \\ \bottomrule
\end{tabular}
}

\end{table}

In Section~\ref{sec:nlp-on-structured-document}, we explain that \tool{} uses \textit{named entity recognition} (NER) to identify data types (labeled as \textit{DATA}) and entities (labeled as \textit{ENTITY}) in the policy text. In this appendix, we provide details on our NER methodology.

\paragraph{Synthetic NER Corpus} Training a new NER model requires a labeled corpus. To avoid the burden of building a dataset from scratch, we programmatically generate a synthetic corpus for NER training. The synthetic corpus is generated from the following two sources.

First, we manually write 30 sentence templates. The sentence templates looks like this: \comptt{We} \texttt{(don't\allowbreak|may\allowbreak|might\allowbreak|...) (share\allowbreak|exchange\allowbreak|...) your \%DATA\% from \%ENTITY\%}, where \texttt{\%DATA\%} and \texttt{\%ENTITY\%} are placeholders for data types and entity names. The regular-expression-like format is reversely expanded to matching strings using exrex~\cite{py-exrex} library. We manually write 90 phrase templates for data types using the same regular-expression-like format, for example: \comptt{(real|first|middle|last|...)? names?}. For entity names, we obtain a list of about 12K entity names by crawling Wikidata~\cite{wikidata} items that are related to companies or products. This includes both proper names (\eg{} company and organization names) and generic category names (\eg{} ``email hosting services''). By using random combinations of sentence templates, data type phrases and entity names, and by expanding the regular expressions, we create a large variety of sentences. For example, a possible sentence created from previous examples is ``We may share your last name with email hosting services.''
As the sentences are created programmatically, the positions of data types and entity names are already known.
We generate 50,000 synthetic sentences as the training set.

Second, we apply knowledge distillation~\cite{knowledge-distillatiion} to transfer knowledge from spaCy's pretrained NER model and our manually defined rule-based NER. Specifically, spaCy's NER model~\cite{spacy-trf-model} is capable to identify company, person and product names, which we map to the \textit{ENTITY} labels. We additionally define rules to label phrases with root words like ``information'', ``data'' and ``identifier'' with  \textit{DATA} labels. Table~\ref{tab:ner-root-word} shows the list of root words used in the rule-based NER. We apply both NER methods to label 200 real-world policies (which is different than our test set in Section~\ref{sec:evaluation}), and include the data in the training set. As this method cannot capture all the entities and data types (otherwise we would simply use it), we mark noun phrases without labels as missing annotations. spaCy's trainer will not penalize nor reward the model for labeling these phrases.

\paragraph{NER Model}
We fine-tune the \textit{distilroberta-base} model~\cite{Sanh2019DistilBERTAD} with the synthetic corpus as the NER model.
To evaluate the model, we randomly selected 200 segments of text, which contain about 620 sentences in total, from our test set of policies (see Section~\ref{sec:evaluation}), and manually label data types and entities in them to establish the ground truth.
On the evaluation corpus, the model achieves 96.1\% precision and 89.4\% recall.
The performance is as good as general-purpose NER models, which have approximately 90\% precision and recall~\cite{spacy-trf-model}.

\subsection{Phrase Normalization}
\label{appendix:phrase-normalization}

In Section~\ref{sec:building-knowledge-graph}, we explain that \tool{} performs phrase normalization using a few strategies. We provide more details on these strategies in this appendix.

To merge phrases with the same meaning (\ie{} synonyms and coreferences) into one node in \poligraph{}, \tool{} maps phrases of data types and entities to their normalized forms.
This process, known as \textit{phrase normalization}~\cite{text-normalization-1}, is necessary for automated analysis, because it allows us to deal with consistent normalized forms instead of many synonymous terms, \eg{} ``contact information'' is the normalized form for ``contact details'' and ``contact info''. \tool{} uses a combination of NLP and manual rules for phrase normalization.
It applies the following strategies to normalize data types and entities.

\paragraph{Standard Terms} We consider data types and entities in our global ontologies (see Section~\ref{sec:ontology}) as standard terms. To facilitate analysis of these data types and entities in Sections~\ref{sec:evaluation} and~\ref{sec:applications}, we write regular expressions to capture variants of synonyms for them.
For example, the regular expression \comptt{contact\b.*\b(information|data|detail|method)} maps ``contact data'', ``contact detail'', ``contact method'', and other synonyms into ``contact information'', which is the normalized form in the global data ontology.

To normalize company names, we build regular expressions from public datasets. We use the DuckDuckGo Tracker Radar dataset~\cite{tracker-radar} and a CrunchBase-based public dataset~\cite{crunchbase-data} to obtain variants of company names, \eg{} ``Alphabet Inc.'' (group company) and ``Firebase'' (product) as alternative names for ``Google'' (the normalized term). We extract n-grams that are uniquely found in each company's list of alternative names to build regular expressions for the company. For example, two-grams ``alphabet inc'' and ``google inc'', and one-grams ``firebase'' and ``google'' are all normalized to ``Google''.

\paragraph{Unspecified Data and Unspecified Third Party}

As a special case of standard terms, we use two special terms ``unspecified data''  and ``unspecified third party'' for data types and entities that have too general meanings.
For data types, if a phrase, after removing stop words, is lemmatized to a single word ``data'', ``information'' or ``detail'', it is normalized to ``unspecified data''.
For entities, if a phrase, after removing stop words, is lemmatized to ``third party'' or similar phrases, it is normalized to ``unspecified third party''.

\paragraph{Other Terms}

At last, if a data type or entity cannot be normalized by abovementioned strategies, \tool{} falls back to strip stop words and take the lemmatized form of the phrase. For example, ``your vehicle records'' is normalized to ``vehicle record''. In most cases, this should be enough to capture variants of the same term caused by word inflections.

\paragraph{Coreferences}
As the coreference annotator has linked each coreference phrase to what it refers to,
\tool{} does not apply normalization to phrases with \COREF{} edges. Instead, it follows the \COREF{} edges to find the referent and use the same normalized form. For example, in Figure~\ref{fig:policy-example}(a), ``this information'' would be normalized to ``geolocation'', same as the phrase ``location'' that it refers to.

\section{Contradiction Analysis in Detail}
\label{appendix:contradiction-analysis}

In Section~\ref{sec:negative-statements}, we presented our methodology of contradiction analysis, and we reported our main findings. In this section, we revisit and provide additional details.

\subsection{Framework Extension}
\label{appendix:contradiction-framework-extension}

In Section~\ref{subsubsec:negative-statements-extensions}, we present extensions to \poligraph{}  for contradiction analysis.
One of the extensions was to include five types of finer-grained actions as subtypes of \COLLECT{} and \NOTCOLLECT{} edges. Table~\ref{tab:verb-to-actions} lists all the verbs and corresponding actions. We note that the modular design of \poligraph{} allows for this list to be easily extended, if so desired.

\begin{table}[t!]
\linespread{0.95}\selectfont\centering
\centering
\caption{Mapping from verbs to subtypes (\ie{} actions) of \COLLECT{} / \NOTCOLLECT{} edges as an extension to \poligraph{}.}
\label{tab:verb-to-actions}
\small
\begin{tabular}{@{}ll@{}}
\toprule
\textbf{Verbs} & \textbf{Example Sentences and Extracted Edges} \\ \midrule

\multirow{2}{*}{\begin{tabular}[c]{@{}p{28mm}@{}}
collect, gather, obtain, receive, get, solicit, acquire, request
\end{tabular}}
& \begin{tabular}[c]{@{}l@{}}
We may collect your device ID.\\
\textit{we}\edge{COLLECT $[collect]$}\textit{device identifier}
\end{tabular} \\
\cmidrule(l){2-2}

& \begin{tabular}[c]{@{}l@{}}
We don't collect your device ID. \\
\textit{we}\edge{NOT\_COLLECT $[collect]$}\textit{device identifier}
\end{tabular}\\ \midrule

\multirow{2}{*}{\begin{tabular}[c]{@{}p{28mm}@{}}
share, exchange, provide, disclose, supply, transmit, release, transfer, submit, give, divulge, pass
\end{tabular}}
& \begin{tabular}[c]{@{}l@{}}
We may share your device ID with Google.\\
\textit{we}\edge{COLLECT $[collect]$}\textit{device identifier} \\
\textit{google}\edge{COLLECT $[be\_shared]$}\textit{device identifier}
\end{tabular} \\
\cmidrule(l){2-2}

& \begin{tabular}[c]{@{}l@{}}
We don't share your device ID with Google. \\
\textit{google}\edge{NOT\_COLLECT $[be\_shared]$}\textit{device identifier}
\end{tabular}\\ \midrule

\multirow{2}{*}{\begin{tabular}[c]{@{}p{28mm}@{}}
sell, rent, lease, trade
\end{tabular}}
& \begin{tabular}[c]{@{}l@{}}
We may sell personal information to Google.\\
\textit{we}\edge{COLLECT $[collect]$}\textit{personal information} \\
\textit{google}\edge{COLLECT $[be\_sold]$}\textit{personal information}
\end{tabular} \\
\cmidrule(l){2-2}

& \begin{tabular}[c]{@{}l@{}}
We don't sell personal information to Google.\\
\textit{google}\edge{NOT\_COLLECT $[be\_sold]$}\scalebox{.9}[1.0]{\textit{personal information}}
\end{tabular}
\\ \midrule

\multirow{2}{*}{\begin{tabular}[c]{@{}p{28mm}@{}}
use, access, process, need,
utilize, analyze,
have / get / gain access to,
make use of
\end{tabular}}
& \begin{tabular}[c]{@{}l@{}}
We may use your device ID.\\
\textit{we}\edge{COLLECT $[use]$}\textit{device identifier}
\end{tabular} \\
\cmidrule(l){2-2}

& \begin{tabular}[c]{@{}l@{}}
We do not use your device ID.\\
\textit{we}\edge{NOT\_COLLECT $[use]$}\textit{device identifier}
\end{tabular}
\\ \midrule

\multirow{2}{*}{\begin{tabular}[c]{@{}p{28mm}@{}}
store, retain, keep, preserve,
record, save, maintain,
log, hold
\end{tabular}}
& \begin{tabular}[c]{@{}l@{}}
We store your device ID on the server.\\
\textit{we}\edge{COLLECT $[store]$}\textit{device identifier}
\end{tabular} \\
\cmidrule(l){2-2}

& \begin{tabular}[c]{@{}l@{}}
We never store your device ID.\\
\textit{we}\edge{NOT\_COLLECT $[store]$}\textit{device identifier}
\end{tabular}
\\ \bottomrule
\end{tabular}

\end{table}

\subsection{Methodology}
\label{appendix:contradiction-methodology}

In Section~\ref{subsubsec:negative-statements-analysis}, we explain how \tool{} determines conflicts between \COLLECT{} and \NOTCOLLECT{} edges.
In contrast, PolicyLint reports contradictions between positive tuples
\textit{$\langle$entity, collect, data type$\rangle$}
and negative tuples
\textit{$\langle$entity, not\_collect, data type$\rangle$}.
While, conceptually, one tuple should correspond to one \COLLECT{} or \NOTCOLLECT{} edge, in practice, it is not always possible to find a one-to-one mapping between PolicyLint tuples and \poligraph{} edges.

First, as PolicyLint and \tool{} implement text preprocessing and phrase normalization in different ways, it is not always possible to match data types, entities or even sentences from both tools. We craft rules to map terms from both tools on a best-effort basis. For example, ``geographical location'' in PolicyLint can match with ``geolocation'', ``precise geolocation'', and ``coarse geolocation'' in \poligraph{}s.

Second, one tuple reported by PolicyLint may come from multiple sentences. As our tool normalizes phrases differently and, more importantly, is extended to capture more parameters, two sentences with the same PolicyLint tuple may result in different edges in \poligraph{}.
In this case, for each pair of contradicting tuples in PolicyLint, \tool{} checks all combinations of \COLLECT{} and \NOTCOLLECT{} edges that match the contradicting tuples, and reports only one pair of edges that have the most conflicting parameters.

On our dataset, PolicyLint reports 2,555 contradictions (excluding ``narrow definitions'' as defined in \cite{andow2019policylint,andow2020actions}) in total, and 1,566 of them are successfully mapped to \poligraph{} edges. In Table~\ref{tab:contradiction-reclassification} in Section~\ref{sec:negative-statements}, we report statistics based on \poligraph{} edges instead of tuples for clarity.

\subsection{Manual Validation of Conflicting Edges}
\label{appendix:contradiction-validation}

In Section~\ref{subsec:validation-of-contraditions}, we briefly discuss the validation of conflicting edges.
We report that, among all the 211 pairs of conflicting edges in our dataset, human readers determines that only 25.1\% (53) pairs are real contradictions.
Apart from 24 pairs that are determined to be NLP errors, we explain the reasons why the remaining conflicting edges are not real contradictions as below.

\paragraph{Additional Contexts} 114 out of 211 pairs of conflicting edges turn out not to be real contradictions because they mention additional contexts that cannot be represented in \poligraph{}, namely:

\begin{compactitem}[$\bullet$]
    \item \textit{Conditions} of data collection. For example, the sentence ``we do not collect personal data when you visit our website'' does not contradict with ``we collect personal data when you submit a request'' due to the different conditions discussed.
    \item \textit{Consent type} of the data collection. For example, many policies declare that ``we do not collect information without your consent''.
    \item \textit{Sources} from which the data are collected. For example, the sentence ``we do not collect personal data from third parties'' does not contradict with ``we collect personal data you provided''. In this case, the sources of data ``third parties'' and ``you'' (the user) are different.
\end{compactitem}

\noindent We note that this is not a fundamental limitation of the \poligraph{} framework, since these additional parameters can be incorporated into \poligraph{} by introducing new edge, or node types and attributes.

\paragraph{Finer Semantics} 20 out of 211 pairs of conflicting edges are not contradictions because the negative and positive sentences discuss subtly different data types, entities, or purposes where the semantic differences are beyond the granularity of \tool{}'s phrase normalization. For example, one policy mentions ``we share your personal information with internal third parties'' but ``not with unaffiliated third parties''. In this case, both phrases of entities are normalized to ``unspecified third party''.
Also note that the new subject annotator is currently able to only identify children as data subjects, while in practice there can be other subjects (\eg{} ``driver'' and ``rider'' in Uber's policy~\cite{uber-privacy-policy}) not handled by the current implementation.

Note that this is not a limitation of the \poligraph{} framework, but of the implementation.
We believe that a more sophisticated approach, likely a model-based approach, to infer the meaning of a phrase, rather than normalizing phrases according to static rules (see Section~\ref{sec:building-knowledge-graph}), is needed to work around such limitations.

\section{\llmtool{} in Detail}

In Section~\ref{sec:llmtool}, we present \llmtool{}, the LLM-based system that we implement to generate \poligraph{}. We provide additional details in this appendix.

\subsection{LLM Prompts for the Annotator Module}
\label{appendix:llm-prompt-annotator}

In \Cref{sec:llm-annotator-module}, we explain how the annotator module in \llmtool{} identifies data collection statements from \privacypolicy{s} and extracts parameters of interest.
We provide the prompts that we used to query the LLM in this appendix.

First, \llmtool{} describes the task using the following system prompt:

{\setstretch{1.0}\scriptsize
\begin{Verbatim}[frame=single,breaklines,breaksymbol={},breakindent=1em,breakautoindent=true]
### Instructions

Analyze the user-provided privacy policy excerpt and extract information about personal data processing.

Return a list of JSON objects, each with the following keys:
- action: List[str] -- List of actions applied to the personal data. For example: "collect", "share", "use".
- data: List[str] -- List of personal data types that are processed. For example: "email address", "mac address", and broader terms like "personal data", "contact info".
- processor: List[str] -- List of entities that process the personal data. For example: "we" (the first party), "our third-party partners", or specific company names.
- recipient: List[str] -- List of entities that receive personal data, when the action involves data transfer. Same examples as for "processor".
- purpose: List[str] -- List of purposes for which the personal data is processed. For example: "authentication", "to provide services".
- context: List[str] -- Other conditions associated with personal data processing. For example: "if you register an account", "when you use our services".
- prohibition: bool -- Specially, if the statement denies or prohibits the stated action (for example, "we DO NOT collect..."), include this key and set it to true.

Notes:
- Ensure that the string values are extracted exactly from the text, preserving the original wording.
- The information to extract may spread across multiple sentences. Make sure to analyze the entire excerpt.
- Omit any of the keys if the corresponding information is not present in the text.
- Only include affirmative and negative statements concerning personal data processing. Ignore other types of statements.
- Return a list of JSON objects, one for each relevant statement found in the excerpt. If there are no relevant statements, simply return an empty list `[]`.

### Examples

Input 1:
> When you create an account, or when you contact us, we may collect a variety of information,
> including your name, mailing address, contact preferences, and credit card information.

Output 1:
[
  {
    "action": ["collect"],
    "processor": ["we"],
    "data": ["name", "mailing address", "contact preferences", "credit card information"],
    "context": ["When you create an account", "when you contact us"]
  }
]

Input 2:
> Here are the types of personal information we collect:
> * Identity Information: such as your user identification number.
> * Contact Information: such as your email address and telephone number.
> We will never share these data with third parties.

Output 2:
[
  {
    "action": ["collect"],
    "processor": ["we"],
    "data": ["Identity Information", "user identification number", "Contact Information", "email address", "telephone number"]
  },
  {
    "action": ["share"],
    "processor": ["We"],
    "recipient": ["third parties"],
    "data": ["Identity Information", "user identification number", "Contact Information", "email address", "telephone number"],
    "prohibition": true
  }
]

Input 3:

> We may share your personal information with CompanyX.
> CompanyX uses your personal information to operate, provide, and improve the products that we offer.
> These purposes include: Purchase and delivery of products.

Output 3:
[
  {
    "action": ["share"],
    "processor": ["We"],
    "recipient": ["CompanyX"],
    "data": ["personal information"],
  },
  {
    "action": ["uses"],
    "processor": ["CompanyX"],
    "data": ["personal information"],
    "purpose": ["to operate, provide, and improve the products that we offer", "Purchase and delivery of products"]
  }
]

Input 4:

> As required by law, we will never disclose sensitive personal information to third parties without your explicit consent.
> When you use third party services, including cloud services and customer service providers, they may share information about that usage with us.

Output 4:
[
  {
    "action": ["disclose"],
    "processor": ["We"],
    "recipient": ["third parties"],
    "data": ["sensitive personal information"],
    "context": ["As required by law", "without your explicit consent"],
    "prohibition": true
  },
  {
    "action": ["share"],
    "processor": ["third party services", "cloud services", "customer service providers"],
    "recipient": ["us"],
    "data": ["information about that usage"],
    "context": ["When you use third party services, including cloud services and customer service providers"]
  }
]

Input 5:
> You have the right to access, update, and correct inaccuracies in your personal information in our custody.
> However, you may not disable certain types of data processing.

Output 5:
[]
\end{Verbatim}
}

The \privacypolicy{} excerpt (\ie{} a text chunk in \Cref{sec:llm-implementation}) is provided in the very first user prompt in the following format:

{\setstretch{1.0}\scriptsize
\begin{Verbatim}[frame=single,breaklines,breaksymbol={},breakindent=1em,breakautoindent=true]
### INPUT

{TEXT CHUNK}
\end{Verbatim}
}

After getting the JSON output, \llmtool{} prompts the LLM to reflect on its answer by asking in the user prompt
``\texttt{Are there still more statements to be added? Answer 'YES' or 'NO'.}''
If the response is yes, the conversation restarts from right after the LLM's JSON output, and the annotator prompts the LLM to continue generation by requesting
``\texttt{Some statements were missed in the last extraction. Please continue.}''
The annotator stops for the current chunk
\begin{inenum}
\item if the LLM answers that there are no more statements to be added, or
\item after 3 rounds of reflection. 
\end{inenum}

\subsection{LLM Prompts for the Data Type Normalization Module}
\label{appendix:llm-prompt-data-type-normalization}

In \Cref{sec:llm-term-norm-implementation}, we explain how \llmtool{} normalizes phrases of data types.
We provide the prompts that we used to query the LLM in this appendix.

First, \llmtool{} describes the task using the following system prompt:

{\setstretch{1.0}\scriptsize
\begin{Verbatim}[frame=single,breaklines,breaksymbol={},breakindent=1em,breakautoindent=true]
### Data Type Ontology

The data type ontology includes the following concepts organized in a hierarchical structure:
- #PersonalIdentifier: Data that uniquely identify an individual.
 - #GovernmentID: Government-issued identification numbers used to uniquely identify individuals.
  - #TaxID: An identification number assigned by a tax authority for taxation purposes, such as a Social Security Number.
  - #PassportNumber: An identification number assigned to an individual's passport.
  - #DriverLicenseNumber: A unique number assigned to an individual's driver's license.
 - #ContactInfo: Data used to contact an individual.
  - #EmailAddress: An electronic mail address used for sending and receiving messages over the internet.
  - #PostalAddress: The physical address where an individual resides or receives mail.
  - #PersonName: The name of an individual, usually including given name(s) and surname(s).
  - #PhoneNumber: A telephone number associated with an individual, used for voice calls or text messaging.
- #DeviceIdentifier: Identifiers associated with electronic devices.
 - #SoftwareIdentifier: Identifiers associated with software applications.
  - #IPAddress: The Internet Protocol address assigned to each device connected to a computer network.
  - #AndroidID: A unique identifier generated for each Android device.
  - #RouterSSID: The Service Set Identifier (SSID), which is the name assigned to a Wi-Fi network by a wireless router.
  - #AdvertisingID: A user-resettable, unique identifier used for advertising purposes on mobile devices.
  - #CookieOrSimilar: Small pieces of data stored on a user's device by websites or applications, such as cookies, E-tags, or pixel tags, mainly used for tracking.
 - #HardwareIdentifier: Identifiers associated with physical hardware devices.
  - #MACAddress: A unique identifier assigned to a network interface controller for communications at the data link layer of a network segment.
  - #IMEI: The International Mobile Equipment Identity, a unique identifier assigned to mobile devices for identification on cellular networks.
  - #SerialNumber: A unique code assigned by the manufacturer to identify a specific device.
- #PersonalCharacteristics: Data about an individual's physical or demographic characteristics.
 - #ProtectedClass: Data relating to legally protected personal characteristics.
  - #Gender: Data about an individual's gender or gender identity.
  - #Ethnicity: Data about an individual's race or ethnic background.
  - #DateOfBirth: The date on which an individual was born.
  - #Age: An individual's age or age range.
 - #Biometric: Biological characteristics used for individual identification.
  - #Voiceprint: A digital representation of an individual's unique vocal characteristics.
  - #Fingerprint: The unique patterns of friction ridges on an individual's finger.
- #Location: Data indicating the geographical location of an individual or device.
 - #CoarseLocation: General location data, such as city or postal code, providing approximate location.
 - #PreciseLocation: Exact geographical coordinates, such as GPS latitude and longitude, providing specific location details.
- #InternetActivity: Data related to an individual's use of the internet and applications.
 - #ApplicationInstalled: Data about software applications installed on a device, such as names, versions, and installation dates.
 - #BrowsingHistory: Records of websites visited by an individual, including URLs and timestamps.
 - #SearchHistory: Records of search queries entered by an individual into search engines.

Below are a few more broader categories of personal data types:
- #DeviceInfo: Data about electronic devices, subsuming #DeviceIdentifier, #Location, and #InternetActivity.
- #PersonalData: Data about individuals, subsuming #PersonalIdentifier, #PersonalCharacteristics, #Location, and #InternetActivity.
- #Identifier: Data that uniquely identify entities, subsuming #PersonalIdentifier and #DeviceIdentifier.

### Instructions

A named entity recognition (NER) model has identified phrases that refer to personal data types in the user-provided privacy policy excerpt.
Below is an example input. The phrases enclosed in braces ({}) are the identified data types:
> We may collect certain {personal information}, including but not limited to {your gender}, {birth date}, and {contact details}.
> We may collect {non-sensitive information} such as {your IP address}, {session ID} and device event log.
> We use {the information we collected} for {advertising}.
> We do not collect {sensitive personal information}. If we accidentally collect {such information}, we will delete it immediately.

Your task is to map these phrases to standard concepts of personal data as defined in the Data Type Ontology.

Return a JSON object that include each of the identified phrases as a key, and the value is an object with one or more of the following keys:
- "concept": if the phrase is synonymous to a standard concept in the ontology, return the corresponding concept.
- "referents": if the phrase subsumes or refers to other phrases in the text, return a list of these phrases.
- "other": if the phrase does not match any standard concept or refer to other phrases, set this key to true.
- Specially, the NER model may have identified irrelevant phrases that are not personal data types. Return a null value for these phrases.

For the previous example, a possible output is:
{
  "certain personal information": {"concept": "#PersonalData", "referents": ["your gender", "birth date", "contact details"]},
  "your gender": {"concept": "#Gender"},
  "birth date": {"concept": "#DateOfBirth"},
  "contact details": {"concept": "#ContactInfo"},
  "non-sensitive information": {"referents": ["your IP address", "session ID", "device event log"]},
  "your IP address": {"concept": "#IPAddress"},
  "session ID": {"other": true},
  "device event log": {"other": true},
  "the information we collected": {"referents": ["personal information", "non-sensitive information"]},
  "advertising": null,
  "sensitive personal information": {"other": true},
  "such information": {"referents": ["sensitive personal information"]}
}

Notes:
- The NER model may have missed some relevant phrases. If you identify any missing personal data types, you can add them to the output with the appropriate mapping (such as `device event log` in the example).
- Only include confident concept mappings. If you are not confident, prefer to mark the phrase as "other" so the next module can double-check these phrases.
- Do not map phrases that do not refer to personal data types (such as `advertising` in the example).
\end{Verbatim}
}

The \privacypolicy{} excerpt is provided in the very first user prompt, and the data type phrases to normalize are quoted in braces, for example:

{\setstretch{1.0}\scriptsize
\begin{Verbatim}[frame=single,breaklines,breaksymbol={},breakindent=1em,breakautoindent=true]
### INPUT

We collect {contact details}, such as {your name}, {e-mail address}, ...
\end{Verbatim}
}

The reflection is already done in the main prompt (\ie{} allowing the LLM to include new data types and discard irrelevant phrases). After a round output, the module continue to the next round by repeating the same system prompt and updated phrase list.
It stops after 3 rounds or the phrase list is not updated.

\subsection{LLM Prompts for the Entity Normalization Module}
\label{appendix:llm-prompt-entity-normalization}

In \Cref{sec:llm-term-norm-implementation}, we explain how \llmtool{} classifies entities into first- and third- parties (and additionally, ``other'' entities).
We provide the prompts that we used to query the LLM in this appendix.

{\setstretch{1.0}\scriptsize
\begin{Verbatim}[frame=single,breaklines,breaksymbol={},breakindent=1em,breakautoindent=true]
### Instructions

You will be presented with a privacy policy excerpt and a list of entity names found in the text.

Your task is to classify each entity as one of:
- "first party", i.e., "we" in the privacy policy;
- "second party", i.e., "you" or "the user" in the privacy policy;
- "third party", i.e., any other data processors or recipients;
- "other", i.e., irrelevant phrases wrongly labeled as entities.

Return the classification result as a JSON object with the entity name as the key and the classification as the value, e.g.:
{
    "Company Foo Ltd": "first party",
    "our service": "first party",
    "you": "second party",
    "children under 13": "second party",
    "our affiliated companies": "first party",
    "Bar Media LLC": "third party",
    "Company Baz Inc": "third party",
    "third-party services": "third party",
    "advertisement networks": "third party",
    "personal data": "other"
}

### Privacy Policy Excerpt

{...}

### Entities

{...}
\end{Verbatim}
}

\subsection{LLM Prompts for the Naive LLM Privacy Policy Analyzer}
\label{appendix:llm-prompt-naive-analyzer}

In \Cref{subsec:llmpg-comparison}, we implement a naive LLM privacy policy analyzer to compare with \llmtool{}. We use the following system prompts to query the LLM:

{\setstretch{1.0}\scriptsize
\begin{Verbatim}[frame=single,breaklines,breaksymbol={},breakindent=1em,breakautoindent=true]
  Analyze the user-provided privacy policy excerpt and determine if the following personal information types are collected, shared, or used:

  - "mac address": A unique identifier assigned to a network interface controller for communications at the data link layer of a network segment.
  - "router ssid": The Service Set Identifier (SSID), which is the name assigned to a Wi-Fi network by a wireless router.
  - "android id": A unique identifier generated for each Android device.
  - "imei": The International Mobile Equipment Identity, a unique identifier assigned to mobile devices for identification on cellular networks.
  - "advertising identifier": A user-resettable, unique identifier used for advertising purposes on mobile devices.
  - "serial number": A unique code assigned by the manufacturer to identify a specific device.
  - "email address": An electronic mail address used for sending and receiving messages over the internet.
  - "phone number": A telephone number associated with an individual, used for voice calls or text messaging.
  - "person name": The name of an individual, usually including given name(s) and surname(s).
  - "geographical location": Data indicating the geographical location of an individual or device.
  
  For each type of personal information, specify if it is collected by, used by, or shared with:
  (1) the first party, i.e., the developer or the owner of the service; and/or
  (2) third parties, i.e., any external service providers that may process personal information.
  
  Return the results as a JSON object, with the data types as keys, and a list of strings as values indicating whether the data is handled by the first party, third parties, or both. For example:
  
  {
    "mac address": ["first party"],
    "advertising identifier": ["third party"],
    "email address": ["first party", "third party"],
    "phone number": ["first party", "third party"],
    "geographical location": ["first party", "third party"]
  }
  
  Notes:
  - Do not overly interpret the text. Only include data types explicitly mentioned in the privacy policy excerpt.
  - The input exerpt may contain no relevant information. In such cases, return an empty object.
\end{Verbatim}
}

\subsubsection{\llmtool{} Component Evaluation}
\label{appendix:llm-component-eval}

In \Cref{sec:llmpg-evaluation}, we evaluate the performance of \llmtool{}.
We provide additional component evaluation results of \llmtool{} in this appendix.

\paragraph{Annotator}
\llmtool{} annotator extracts 13,586 data collection statements in total.
We analyze the extracted data collection statements to see how frequent hallucination and coverage errors are fixed.
Recall that the annotator discard statements that do not appear in the text, and performs three rounds of reflection to improve the coverage (see \Cref{sec:llmpg-design-decisions}).
\Cref{tab:annotator-rounds} shows the number of valid and discarded statements reported by the annotator in each round.
On the one hand, 7.4\% of extracted statements are not valid, likely caused by hallucination.
A common issue is that the LLM copies the few-shot-learning examples in our prompt into the output.
On the other hand, reflection helps discover 2.2\% more of statements.

\begin{table}[t!]
\centering
\caption{Number of new valid and discarded statements reported by the annotator in each round of reflection.}
\label{tab:annotator-rounds}
\centering
\begin{tabular}{rrr}
\toprule
            & \# valid & \# discarded \\ \midrule
    Round 1 &   13,291 (93.1\%)  &    983 (\hphantom{0}6.9\%) \\
    Round 2 &      265 (74.2\%)  &    93 (26.0\%) \\
    Round 3 &       30 (76.9\%)  &     9 (23.1\%) \\
{\em total} &   13,586 (92.6\%)  & 1,085 (\hphantom{0}7.4\%)  \\
\bottomrule
\end{tabular}
\end{table}

\paragraph{Term Normalization}
To evaluate the term normalization modules, we randomly sample 200 data types, entities and purpose phrases, and manually verify if they are mapped correctly.
We exclude data types that are classified as \path{#Other}, and entities that do not represent first or third parties (\eg{} the user) from the statistics.
The data type normalization module achieves 91.0\% precision in mapping data types to the ontology nodes.
The entity normalization module achieves 98.5\% accuracy in distinguishing first- and third-party entities.
The purpose normalization module achieves 97.0\% and 93.0\% macro-averaged precision and recall, respectively, for the multi-label multi-class classification task.

\end{document}